\documentclass{llncs}
\usepackage{amsmath}
\usepackage{amssymb}
\usepackage{url}
\usepackage{bussproofs}
\usepackage{cite}
\usepackage[ruled,vlined]{algorithm2e}
\usepackage{footmisc}
\usepackage{tikz}

\newcommand{\n}[1]{\overline{#1}}

\newcommand{\nin}{\noindent}
\newcommand{\lb}{\linebreak}
\newcommand{\UFIDL}{\ensuremath{\texttt{QF\_UFIDL}}\xspace}

\newcommand{\IDL}{\ensuremath{\mathcal{IDL}}\xspace}

\newcommand{\LRA}{\ensuremath{\mathcal{LRA}}\xspace}
\newcommand{\EUF}{\ensuremath{\mathcal{EUF}}\xspace}
\newcommand{\LIA}{\ensuremath{\mathcal{LIA}}\xspace}
\newcommand{\AX}{\ensuremath{\mathcal{AX}}\xspace}
\newcommand{\T}{\ensuremath{\mathcal{T}}\xspace}

\newcommand{\opensmt}{{\sc OpenSMT}\xspace}
\newcommand{\periplo}{{\sc PeRIPLO}\xspace}

\newtheorem{thm}{Theorem}

\begin{document}

\title{Resolution Proof Transformation for Compression and Interpolation}



\author{S.F. Rollini\inst{1} \and R. Bruttomesso\inst{2} \and N. Sharygina \inst{1} \and A. Tsitovich \inst{3}}

\institute{
University of Lugano, Switzerland\\
\email{\{simone.fulvio.rollini, natasha.sharygina\}@usi.ch}
\and
 Atrenta Advanced R\&D of Grenoble, France \\
              \email{roberto@atrenta.com}
\and
Phonak, Switzerland \\
            \email{aliaksei.tsitovich@phonak.com}
}

\maketitle

\begin{abstract}
Verification methods based on SAT, SMT, and Theorem Proving 
often rely on proofs of unsatisfiability as a powerful tool
to extract information in order to reduce the overall effort. 
For example a proof may be traversed to identify a 
minimal reason that led to unsatisfiability, for 
computing abstractions, or for deriving Craig interpolants.
In this paper we focus on two important aspects that concern
efficient handling of proofs of unsatisfiability:
compression and manipulation. First of all, since the proof size
can be very large in general (exponential in the size of the input 
problem), it is indeed beneficial to adopt techniques to compress it 
for further processing. Secondly, proofs can be manipulated as a flexible 
preprocessing step in preparation for interpolant computation. Both these techniques are
implemented in a framework that makes use of local rewriting rules
to transform the proofs. We show that a careful use of the rules, combined
with existing algorithms, can result in an effective simplification
of the original proofs. We have evaluated several heuristics on 
a wide range of unsatisfiable problems deriving from SAT and SMT test cases.

\end{abstract}
%

%
%

\section{Introduction}
\label{sec:intro}

Symbolic verification methods rely on a representation of the state space
as a set of formulae, which are manipulated by formal engines such as
SAT- and SMT-solvers. For example Bounded Model Checking~\cite{BCC+03} represents
an execution trace leading to a state violating a property as a propositional formula such
that the state is reachable if and only if the formula is satisfiable.
When the formula is satisfiable, it is possible to infer a counterexample
from the model reported by the solver, showing a path that reaches a violating state. 
When the formula is unsatisfiable, it is instead possible to extract information 
that better explains the reason why the violating states are unreachable. For instance
this can be useful to derive an abstraction of a set of states as it is
done in interpolation-based model checking~\cite{McM03} (to abstract the initial states) 
or IC3~\cite{Bra11} (to derive a minimal set of clauses to put in a frame).

In this paper we describe a set of techniques that allow efficient
manipulation of a propositional proof of unsatisfiability, the by-product of an
unsatisfiable run of a state-of-the-art solver that may be used to obtain abstractions in the
applications mentioned above. In particular we focus on two important aspects:
compression of a proof of unsatisfiability, and rewriting to facilitate the computation
of interpolants. These approaches are both realized by means of a set of local 
rewriting rules that enable proof restructuring and compression. 

\subsection{Structure of the Paper}

The paper is organized as follows. \S\ref{sec:smt} 
recalls some notions about SAT, SMT and resolution proofs.
\S\ref{sec:tra_fra} introduces a proof transformation framework consisting of 
a set of local rewriting rules and discusses its soundness.
\S\ref{sec:compr} addresses the problem of compressing resolution proofs,
proposing a collection of algorithms based on the transformation framework.
It compares them against existing compression techniques and provides experimental results
of running different tools over SMT and SAT benchmarks.
\S\ref{sec:reordtrasf} presents basic notions about interpolation in first order theories and discusses
some limitations of state-of-the-art interpolation algorithms.
It then proposes an application of the transformation framework aimed at reordering
resolution proofs, in such a way that interpolation is made possible. The approach is demonstrated to be theoretically sound
and experiments are provided to show that it is also practically efficient. An algorithm is also provided to reorder resolution steps in a propositional proof
to guarantee the generation of interpolants in conjunctive and disjunctive normal form.
\S\ref{sec:heuri} discusses some of the heuristics adopted in the application of rules
by the transformation framework, with reference to \S\ref{sec:compr} and \S\ref{sec:reordtrasf}.
\S\ref{sec:relwork} reviews the existing literature on proof manipulation; \S\ref{sec:concl} draws the conclusions.

\subsection{Improvement over Previous Own Work}

The present work builds upon and extends~\cite{BRST10} and~\cite{RBS10} in a number of ways:
$(i)$ it gives a unified and richer description of the Local Transformation Framework and of the set of rewriting rules
on which this is based (\S\ref{sec:tra_fra}); $(ii)$ it provides a more thorough comparison between the notions of resolution proof trees and DAGs,
describing the application and the effect of the rules (\S\ref{sec:smt}, \S\ref{sec:tra_fra});
$(iii)$ it gives a proof of correctness of the rules and of the SubsumptionPropagation algorithm presented in~\cite{BRST10} (\S\ref{sec:tra_fra});
$(iv)$ it proposes a new meta-algorithm for proof transformation, TransformAndReconstruct, discussing its soundness
and how it can be instantiated to concrete algorithms with different goals (\S\ref{sec:tra_fra});
$(v)$ two new compression algorithms, PushdownUnits and StructuralHashing, are proposed and discussed, 
as well as their combination with the algorithms in~\cite{BRST10} and~\cite{RBS10} (\S\ref{sec:compr}); 
$(vi)$ a thorough evaluation of previous and novel algorithms is carried out on a set of purely propositional 
benchmarks from the literature (\S\ref{sec:compr});
$(vii)$ in the context of interpolation we illustrate an application of the Local Transformation Framework to reorder 
the pivots in a proof so as to guarantee the generation of interpolants in CNF and DNF (\S\ref{sec:reordtrasf});
$(viii)$ a description of the heuristics adopted in the application of the rewriting rules has been added, 
with reference both to compression and to transformation for interpolation (\S\ref{sec:heuri}).

\section{Background}
\label{sec:smt}

The context of this paper is first order logic. 
We assume countable sets of individual variables ($x,y,z$), function ($f,g$) and predicate ($P,Q$) symbols.
A function symbol of 0-arity is called a constant ($a,b,c$), while a predicate symbol of 0-arity 
corresponds to a propositional variable ($o,p,q,r$). A term is built from function symbols and individual variables ($f(c,g(x))$);
an atom is built from predicate symbols and terms ($P(x,f(c))$). 
A literal ($s,t$) is either an atom (having \emph{positive  polarity}) or its negation (having negative polarity).
A formula ($\phi,\psi$) is built from atoms and connectives; we are only interested here in quantifier-free formulae.
A sentence (or ground formula) is a formula without free variables.
A \emph{clause} $C$ is a finite disjunction of
literals; a formula in conjunctive normal form (\emph{CNF}) is a
finite conjunction of clauses. 
The empty clause, which represents unsatisfiability, is denoted by $\bot$.
We write clauses as lists of literals and sub-clauses,
omitting the ``$\vee$'' symbol, as for instance $p \n{q} D$ (an overline denotes negation).
We use $C_1 \subseteq C_2$ to indicate that $C_1$ \emph{subsumes} $C_2$, 
that is the set of literals $C_1$ is a subset of the set of 
literals $C_2$. 
Also we assume that clauses do not contain duplicated literals or
both the occurrence of a literal and its negation. Finally, we use $v(s)$ to
denote the variable associated with a literal $s$.

A \emph{SAT-solver} is a decision procedure that solves the propositional satisfiability problem;
most successful state-of-the-art solvers rely on variants of the \emph{DPLL} algorithm, as the \emph{conflict-driven clause-learning (CDCL)}~\cite{SS96,bayardo1997using},
which are based on the \emph{resolution inference system}~\cite{Gom08}. 
A first order \emph{theory} \T is a collection of sentences; 
we call \emph{SMT(\T)} the problem of deciding the satisfiability of a formula w.r.t. a theory \T.
A \emph{theory solver} is an algorithm that decides whether a conjunction of ground literals is satisfiable in \T.
If the conjunction is unsatisfiable in \T, then its negation is valid and is called a \emph{\T-lemma}:
intuitively, \T-lemmata are formulae that encode facts valid in the theory \T.
An \emph{SMT(\T)-solver} is a procedure to solve SMT(\T); in particular, a \emph{lazy} solver integrates
a theory solver with a CDCL SAT-solver~\cite{Seb07}.

\subsection{The Resolution System}
The \emph{resolution system} is an inference system based on a single inference rule, called {\em resolution rule}:
\begin{prooftree}
\AxiomC{ $p \, D$ }
\AxiomC{ $\n{p} E$ }
\RightLabel{$p$}
\BinaryInfC{ $D E$ }
\end{prooftree}

\nin{}Clauses $p D$ and $\n{p} E$ are the {\em antecedents},
$D E$ is the {\em resolvent} and $p$ is the {\em pivot} variable. 
We also represent a \emph{resolution step} (an application of the resolution rule) as $DE=Res_p(pD,\n{p}E)$.


SAT- and SMT-solvers can be instrumented to generate, for unsatisfiable formulae, a certificate of unsatisfiability in the form of a \emph{proof of unsatisfiability}
or \emph{refutation}.
It is straightforward
to instruct a state-of-the-art CDCL solver to return
proofs: a resolution proof, in particular, can be derived by 
logging the inference steps performed during conflict analysis~\cite{ZM03b}.

Throughout the paper we shall use the notions of {\em resolution proof tree} and \emph{resolution proof DAG}.
\begin{definition}[Resolution Proof Tree]
\label{def:rpt}
A {\em resolution proof tree} of a clause $C$ from a set of clauses
$\mathbb{C}$ is a tree such that:
\begin{enumerate}
\item Each node $n$ is labeled by a clause $C(n)$.
\item If $n$ is a leaf, $C(n) \in \mathbb{C}$  
\item  The root is a node $n$ s.t. $C(n)=C$.
\item An inner node $n$ has pivot $piv(n)$ and exactly two parents $n^+,n^-$ s.t. $C(n)=Res_{piv(n)}(C(n^+),C(n^-)))$.
$C(n^+)$ and $C(n^-)$ respectively contain the positive and negative occurrence of the pivot.
\item Each non-root node has exactly one child.
\end{enumerate}
\end{definition}
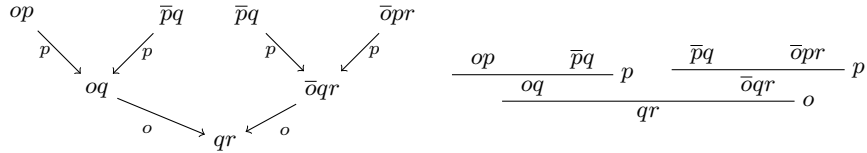
\begin{figure}[!ht]
\centering
\begin{minipage}{.46\textwidth}
\centering
\begin{tikzpicture}[->,every node/.style={font=\sffamily\small}]
  \node (1) {$op$};
  \node (2) [right of=1] {};
  \node (3) [right of=2] {$\n{p}q$};
  \node (5) [right of=3] {$\n{p}q$ };
  \node (6) [right of=5] {};
  \node (7) [right of=6] {$\n{o}pr$};
  \node (8) [below of=2] {$oq$};
  \node (9) [right of=8] {};
  \node (10) [below of=6] {$\n{o}qr$};
  \node (11) [below right of=9] {$qr$};

  \path[every node/.style={font=\sffamily\scriptsize}]
    (1) edge node [left] {$p$} (8)
    (3) edge node [right] {$p$} (8)
    (5) edge node [left] {$p$} (10)
    (7) edge node [right] {$p$} (10)
    (8) edge node [below left] {$o$} (11)
    (10) edge node [below right] {$o$} (11);
\end{tikzpicture}
\end{minipage}
\hspace{1mm}
\begin{minipage}{.46\textwidth}
\centering
\begin{prooftree}
\AxiomC{ $op$ }
\AxiomC{ $\n{p}q$ }
\RightLabel{$p$}
\BinaryInfC{ $oq$ }
\AxiomC{ $\n{p}q$ }
\AxiomC{ $\n{o}pr$ }
\RightLabel{$p$}
\BinaryInfC{ $\n{o}qr$ }
\RightLabel{$o$}
\BinaryInfC{ $qr$ }
\end{prooftree}
\end{minipage}
\caption{Resolution proof tree.}
\label{fig:rpt}
\end{figure}
In the following, we equivalently use  a graph-based representation (left) or
an inference rule-based representation (right).

In real-world applications proofs are rarely generated or stored as
trees; for instance proofs produced by CDCL  solvers are represented
as DAGs (Directed Acyclic Graph).
We therefore introduce the following notion of resolution proof, 
which is more suitable for describing the 
graph-based transformation algorithms illustrated in this paper.

\begin{definition}[Resolution Proof DAG]
\label{def:proof}
A {\em resolution  proof DAG } of a clause $C$ from a set of clauses
$\mathbb{C}$ is a Directed Acyclic Graph such that:
\begin{enumerate}
\item[1.]$\!\!$- 4. hold as in Def.~\ref{def:rpt}.
\item[5.] Each non-root node has one or more children.
\end{enumerate}
Resolution proof DAGs extend the notion of resolution proof trees by allowing a node to participate as antecedent
in multiple resolution steps.
\end{definition}
\begin{figure}[!ht]
\centering
\begin{minipage}{.46\textwidth}
\centering
\begin{tikzpicture}[->,every node/.style={font=\sffamily\small}]
  \node (1) {$op$};
  \node (2) [right of=1] {};
  \node (3) [right of=2] {$\n{p}q$};
  \node (6) [right of=3] {};
  \node (7) [right of=6] {$\n{o}pr$};
  \node (8) [below of=2] {$oq$};
  \node (9) [below of=3] {};
  \node (10) [below of=6] {$\n{o}qr$};
  \node (11) [below of=9] {$qr$};

  \path[every node/.style={font=\sffamily\scriptsize}]
    (1) edge node [left] {$p$} (8)
    (3) edge node [right] {$p$} (8)
    (3) edge node [left] {$p$} (10)
    (7) edge node [right] {$p$} (10)
    (8) edge node [left] {$o$} (11)
    (10) edge node [right] {$o$} (11);
\end{tikzpicture}
\end{minipage}
\caption{Resolution proof DAG.}
\label{fig:rpd}
\end{figure}
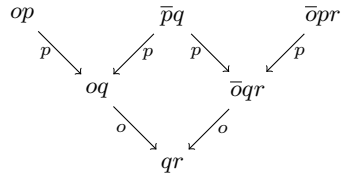
We identify a node $n$ with its clause $C(n)$ whenever convenient; in general, different nodes can be labeled by the same clause, that is $C(n_i)=C(n_j)$
for $i\neq j$.
A proof $P$ is a \emph{refutation} 
if $C=\bot$. A \emph{subproof} $P'$, with \emph{subroot} $n$, 
of a proof $P$ is the subtree that derives $C(n)$ from
a subset of clauses that label leaves of $P$; when referring to $P$ and its root compared to $P'$,
we call $P$ \emph{global proof} and its root \emph{global root}.

It is always possible to turn a resolution proof tree into a resolution proof DAG,
by merging two or more nodes labeled by a same clause into a single node, which inherits 
the children of the merged nodes. 
On the other hand, a resolution proof DAG  can
be ``unrolled'' into a resolution proof tree,
possibly at exponential cost: it is in fact sufficient to traverse the DAG bottom-up,
duplicating nodes with multiple children so that each node is left with at most one child. 

Similarly to~\cite{BFHSS08}, we  distinguish between a \emph{legal} and an \emph{illegal} proof;
an illegal proof is a proof which 
has undergone transformations in such a way that some clauses might not be the 
resolvents of their antecedents anymore. In this paper however an illegal
proof represents an intermediate transformation step in an algorithm, and the proof 
can always be {\em reconstructed} into a legal one, as explained in the next 
sections.

In the following, we will consider refutations as obtained by means of modern  CDCL SAT-solvers and lazy SMT-solvers,
involving both propositional and theory atoms.
Whenever the theory content is not relevant to the problem at hand, it is convenient to represent each theory
atom with a new propositional variable called its \emph{propositional abstraction}: for example an atom $x + y < 1$ will
be represented by a certain variable $q$.

\subsection{Resolution Proofs in Verification}

Resolution proofs
find application in many verification techniques.
For instance, Amla and McMillan's~\cite{AM03} method for
automatic abstraction uses proofs of unsatisfiability derived 
from SAT-based bounded model checking as a guide for choosing 
an abstraction for unbounded model checking. Proofs can be used
as justifications of specifications of inconsistency in various 
industrial applications (e.g., product configuration or declarative modeling\cite{SKK03,SSJSM03}).
In the context of proof-carrying code~\cite{N97} a system can verify a property about an application 
exploiting a proof provided with the application executable code.
SAT-solvers and SMT-solvers can be integrated into interactive theorem provers as automated engines to produce proofs,
that can be later replayed and verified within the provers~\cite{A08,WA09,FMMNT06}.
An unsatisfiable core, that is an inconsistent subset of clauses, can be extracted from a proof, 
to be exploited for example during the refinement phase in model checking~\cite{AM03,GLSM05}.
Another noteworthy application is in the framework of
interpolation-based model checking, where interpolants are generated from proofs based on their structure
and content~\cite{McM03,McM04a,McM04b,HJM+04}.

\section{The Local Transformation Framework}
\label{sec:tra_fra}

This section introduces a proof transformation framework
based on local rewriting rules. 
We start by assuming a resolution proof tree, and then extend the
discussion to resolution proof DAGs. All results related to proofs
hold in particular for refutations.

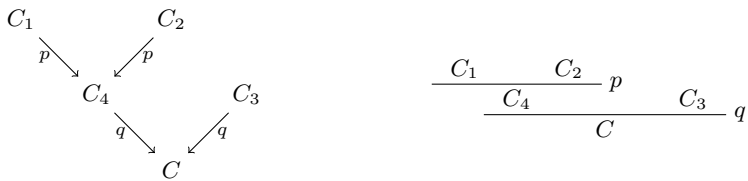
\begin{figure}[h]
\centering
\begin{minipage}{.4\textwidth}
\begin{tikzpicture}[->,every node/.style={font=\sffamily\small}]
  \node (1) {$C_1$};
  \node (2) [right of=1] {};
  \node (3) [right of=2] {$C_2$};
  \node (4) [below of=2] {$C_4$};
  \node (5) [right of=4] {};
  \node (6) [right of=5] {$C_3$};
  \node (7) [below of=5] { $C$};
      \node (8) [below of=1] {};
  \node (9) [below of=8]{};

  \path[every node/.style={font=\sffamily\scriptsize}]
    (1) edge node [left] {$p$} (4)
    (3) edge node [right] {$p$} (4)
    (4) edge node [left] {$q$} (7)
    (6) edge node [right] {$q$} (7); 
\end{tikzpicture}
\end{minipage}
\hspace{2mm}
\begin{minipage}{.4\textwidth}
\begin{prooftree}
\AxiomC{ $C_1$ }
\AxiomC{ $C_2$ }
\RightLabel{$p$}
\BinaryInfC{ $C_4$ }
\AxiomC{ $C_3$ }
\RightLabel{$q$}
\BinaryInfC{ $C$ }
\end{prooftree}
\end{minipage}
\caption{Rule context.}
\label{fig:cont}
\end{figure}

The framework is built on a set of rewriting rules that transform
a subproof with root $C$ into one whose subroot $C'$ is logically equivalent or stronger than $C$ (that is, $C'\implies C$). 
Each rewriting
rule is defined to match a particular {\em context}, identified
by two consecutive resolution steps (see Fig.~\ref{fig:cont}).
A context involves two pivots $p$ and $q$ and five clauses $C_1, C_2, C_4, C_3, C$;
we call $C$ the \emph{context root}; the subproof rooted in $C$ is the \emph{context subproof}.
Clearly $p$ is contained in $C_1$ and $C_2$ (with opposite polarity), and
$q$ is contained in $C_4$ and $C_3$ (again with opposite polarity); $q$ must be contained in $C_1 \cup C_2$.

A clause $C$ might be the root of two different contexts, depending on whether $C_1$ and $C_2$
are taken as the antecedents of either of the two antecedents of $C$; in that case, 
to distinguish among them we talk about \emph{left}
and \emph{right context}.

\begin{figure}[!htp]
\centering
\renewcommand{\arraystretch}{0.34}
\begin{tabular}{|p{0.33\textwidth} p{0.03\textwidth} p{0.53\textwidth}|}
\hline
& & \\
\multicolumn{3}{|c|}{\scriptsize{ $S1$: $s \notin C_3$, $t \in C_2$}} \\
& & \\
\begin{minipage}{0.33\textwidth}
\scalebox{0.85}{
\begin{tikzpicture}[->,every node/.style={font=\sffamily\small}]
  \node (1) {$C_1 : s t D$};
  \node (2) [right of=1] {};
  \node (3) [right of=2] {$C_2: \n{s} t E$};
  \node (4) [below of=2] {$C_4 : t D E$};
  \node (5) [right of=4] {};
  \node (6) [right of=5] {$C_3: \n{t} F$};
  \node (7) [below of=5] { $C : D E F$};

  \path[every node/.style={font=\sffamily\scriptsize}]
    (1) edge node [left] {$v(s)$} (4)
    (3) edge node [right] {$v(s)$} (4)
    (4) edge node [left] {$v(t)$} (7)
    (6) edge node [right] {$v(t)$} (7);
\end{tikzpicture}
}
\end{minipage}
&
$\Rrightarrow$ 
&
\begin{minipage}{0.53\textwidth}
\scalebox{0.85}{
\begin{tikzpicture}[->,every node/.style={font=\sffamily\small}]
  \node (1) {$C_1: s t D$};
  \node (2) [right of=1] {};
  \node (3) [right of=2] {$C_3: \n{t} F$};
  \node (4) [right of=3] {};
  \node (5) [right of=4] {$C_3: \n{t} F$ };
  \node (6) [right of=5] {};
  \node (7) [right of=6] {$C_2: \n{s} t E$};
  \node (8) [below of=2] {$C_4': s D F$};
  \node (9) [below of=4] {};
  \node (10) [below of=6] {$C_4'': \n{s} E F$};
  \node (11) [below of=9] {$C: D E F$};

  \path[every node/.style={font=\sffamily\scriptsize}]
    (1) edge node [left] {$v(t)$} (8)
    (3) edge node [right] {$v(t)$} (8)
    (5) edge node [left] {$v(t)$} (10)
    (7) edge node [right] {$v(t)$} (10)
    (8) edge node [left] {$v(s)$} (11)
    (10) edge node [right] {$v(s)$} (11);
\end{tikzpicture}
}
\end{minipage}\\
& & \\
\hline
& & \\
\multicolumn{3}{|c|}{\scriptsize{ $S1'$: $s \notin C_3$, $t \in C_2$}} \\
& & \\
\begin{minipage}{0.33\textwidth}
\scalebox{0.85}{
\begin{tikzpicture}[->,every node/.style={font=\sffamily\small}]
  \node (1) {$C_1 : s t D$};
  \node (2) [right of=1] {};
  \node (3) [right of=2] {$C_2: \n{s} t E$};
  \node (4) [below of=2] {$C_4 : t D E$};
  \node (5) [right of=4] {};
  \node (6) [right of=5] {$C_3: \n{t} F$};
  \node (7) [below of=5] { $C : D E F$};

  \path[every node/.style={font=\sffamily\scriptsize}]
    (1) edge node [left] {$v(s)$} (4)
    (3) edge node [right] {$v(s)$} (4)
    (4) edge node [left] {$v(t)$} (7)
    (6) edge node [right] {$v(t)$} (7);
\end{tikzpicture}
}
\end{minipage}
& $\Lleftarrow$ &
\begin{minipage}{0.53\textwidth}
\scalebox{0.85}{
\begin{tikzpicture}[->,every node/.style={font=\sffamily\small}]
  \node (1) {$C_1: s t D$};
  \node (2) [right of=1] {};
  \node (3) [right of=2] {$C_3: \n{t} F$};
  \node (4) [right of=3] {};
  \node (5) [right of=4] {$C_3: \n{t} F$ };
  \node (6) [right of=5] {};
  \node (7) [right of=6] {$C_2: \n{s} t E$};
  \node (8) [below of=2] {$C_4': s D F$};
  \node (9) [below of=4] {};
  \node (10) [below of=6] {$C_4'': \n{s} E F$};
  \node (11) [below of=9] {$C: D E F$};

  \path[every node/.style={font=\sffamily\scriptsize}]
    (1) edge node [left] {$v(t)$} (8)
    (3) edge node [right] {$v(t)$} (8)
    (5) edge node [left] {$v(t)$} (10)
    (7) edge node [right] {$v(t)$} (10)
    (8) edge node [left] {$v(s)$} (11)
    (10) edge node [right] {$v(s)$} (11);
\end{tikzpicture}
}
\end{minipage} \\
& & \\
\hline
& & \\
\multicolumn{3}{|c|}{\scriptsize{ $S2$: $s \notin C_3$, $t \notin C_2$}} \\
& & \\
\begin{minipage}{0.33\textwidth}
\scalebox{0.85}{
\begin{tikzpicture}[->,every node/.style={font=\sffamily\small}]
  \node (1) {$C_1 : s t D$};
  \node (2) [right of=1] {};
  \node (3) [right of=2] {$C_2: \n{s} E$};
  \node (4) [below of=2] {$C_4 : t D E$};
  \node (5) [right of=4] {};
  \node (6) [right of=5] {$C_3: \n{t} F$};
  \node (7) [below of=5] { $C : D E F$};

  \path[every node/.style={font=\sffamily\scriptsize}]
    (1) edge node [left] {$v(s)$} (4)
    (3) edge node [right] {$v(s)$} (4)
    (4) edge node [left] {$v(t)$} (7)
    (6) edge node [right] {$v(t)$} (7);
\end{tikzpicture}
}
\end{minipage}
& $\Rrightarrow$ &
\begin{minipage}{0.53\textwidth}
\scalebox{0.85}{
\begin{tikzpicture}[->,every node/.style={font=\sffamily\small}]
  \node (1) {$C_1 : s t D$};
  \node (2) [right of=1] {};
  \node (3) [right of=2] {$C_3: \n{t} F$};
  \node (4) [below of=2] {$C_4': s D F$};
  \node (5) [right of=4] {};
  \node (6) [right of=5] {$C_2: \n{s} E$};
  \node (7) [below of=5] { $C' : D E F$};

  \path[every node/.style={font=\sffamily\scriptsize}]
    (1) edge node [left] {$v(t)$} (4)
    (3) edge node [right] {$v(t)$} (4)
    (4) edge node [left] {$v(s)$} (7)
    (6) edge node [right] {$v(s)$} (7);
\end{tikzpicture}
}
\end{minipage}\\
& & \\
\hline
& & \\
\multicolumn{3}{|c|}{\scriptsize{ $R1$: $s \in C_3$, $t \in C_2$}} \\
& & \\
\begin{minipage}{0.33\textwidth}
\scalebox{0.85}{
\begin{tikzpicture}[->,every node/.style={font=\sffamily\small}]
  \node (1) {$C_1 : s t D$};
  \node (2) [right of=1] {};
  \node (3) [right of=2] {$C_2: \n{s} t E$};
  \node (4) [below of=2] {$C_4 : t D E$};
  \node (5) [right of=4] {};
  \node (6) [right of=5] {$C_3: s \n{t} F$};
  \node (7) [below of=5] { $C: s D E F$};

  \path[every node/.style={font=\sffamily\scriptsize}]
    (1) edge node [left] {$v(s)$} (4)
    (3) edge node [right] {$v(s)$} (4)
    (4) edge node [left] {$v(t)$} (7)
    (6) edge node [right] {$v(t)$} (7);
\end{tikzpicture}
}
\end{minipage}
& $\Rrightarrow$ &
\begin{minipage}{0.53\textwidth}
\scalebox{0.85}{
\begin{tikzpicture}[->,every node/.style={font=\sffamily\small}]
  \node (1) {$C_1 : s t D$};
  \node (2) [right of=1] {};
  \node (3) [right of=2] {$C_3: s \n{t} F$};
  \node (4) [below of=2] {$C' : s D F$};

  \path[every node/.style={font=\sffamily\scriptsize}]
    (1) edge node [left] {$v(t)$} (4)
    (3) edge node [right] {$v(t)$} (4);
\end{tikzpicture}
}
\end{minipage}\\
& & \\
\hline
& & \\
\multicolumn{3}{|c|}{\scriptsize{ $R2$: $s \in C_3$, $t \notin C_2$}} \\
& & \\
\begin{minipage}{0.33\textwidth}
\scalebox{0.85}{
\begin{tikzpicture}[->,every node/.style={font=\sffamily\small}]
  \node (1) {$C_1 : s t D$};
  \node (2) [right of=1] {};
  \node (3) [right of=2] {$C_2: \n{s} E$};
  \node (4) [below of=2] {$C_4 : t D E$};
  \node (5) [right of=4] {};
  \node (6) [right of=5] {$C_3: s \n{t} F$};
  \node (7) [below of=5] { $C : s D E F$};

  \path[every node/.style={font=\sffamily\scriptsize}]
    (1) edge node [left] {$v(s)$} (4)
    (3) edge node [right] {$v(s)$} (4)
    (4) edge node [left] {$v(t)$} (7)
    (6) edge node [right] {$v(t)$} (7);
\end{tikzpicture}
}
\end{minipage}
& $\Rrightarrow$ &
\begin{minipage}{0.53\textwidth}
\scalebox{0.85}{
\begin{tikzpicture}[->,every node/.style={font=\sffamily\small}]
  \node (1) {$C_1 : s t D$};
  \node (2) [right of=1] {};
  \node (3) [right of=2] {$C_3: s \n{t} F$};
  \node (4) [below of=2] {$C_4': s D F$};
  \node (5) [right of=4] {};
  \node (6) [right of=5] {$C_2: \n{s} E$};
  \node (7) [below of=5] { $C' : D E F$};

  \path[every node/.style={font=\sffamily\scriptsize}]
    (1) edge node [left] {$v(t)$} (4)
    (3) edge node [right] {$v(t)$} (4)
    (4) edge node [left] {$v(s)$} (7)
    (6) edge node [right] {$v(s)$} (7);
\end{tikzpicture}
}
\end{minipage}\\
& & \\
\hline
& & \\
\multicolumn{3}{|c|}{\scriptsize{ $R2'$: $s \in C_3$, $t \notin C_2$}} \\
& & \\
\begin{minipage}{0.33\textwidth}
\scalebox{0.85}{
\begin{tikzpicture}[->,every node/.style={font=\sffamily\small}]
  \node (1) {$C_1 : s t D$};
  \node (2) [right of=1] {};
  \node (3) [right of=2] {$C_2: \n{s} E$};
  \node (4) [below of=2] {$C_4 : t D E$};
  \node (5) [right of=4] {};
  \node (6) [right of=5] {$C_3: s \n{t} F$};
  \node (7) [below of=5] { $C : s D E F$};

  \path[every node/.style={font=\sffamily\scriptsize}]
    (1) edge node [left] {$v(s)$} (4)
    (3) edge node [right] {$v(s)$} (4)
    (4) edge node [left] {$v(t)$} (7)
    (6) edge node [right] {$v(t)$} (7);
\end{tikzpicture}
}
\end{minipage}
& $\Rrightarrow$ &
\begin{minipage}{0.53\textwidth}
\scalebox{0.85}{
\begin{tikzpicture}[->,every node/.style={font=\sffamily\small}]
  \node (1) {$C_1 : s t D$};
  \node (2) [right of=1] {};
  \node (3) [right of=2] {$C_3: s \n{t} F$};
  \node (4) [below of=2] {$C' : s D F$};

  \path[every node/.style={font=\sffamily\scriptsize}]
    (1) edge node [left] {$v(t)$} (4)
    (3) edge node [right] {$v(t)$} (4);
\end{tikzpicture}
}
\end{minipage}\\
& & \\
\hline
& & \\
\multicolumn{3}{|c|}{\scriptsize{ $R3$: $\n{s} \in C_3$, $t \not\in C_2$}} \\
& & \\
\begin{minipage}{0.33\textwidth}
\scalebox{0.85}{
\begin{tikzpicture}[->,every node/.style={font=\sffamily\small}]
  \node (1) {$C_1 : s t D$};
  \node (2) [right of=1] {};
  \node (3) [right of=2] {$C_2: \n{s} E$};
  \node (4) [below of=2] {$C_4 : t D E$};
  \node (5) [right of=4] {};
  \node (6) [right of=5] {$C_3: \n{s} \n{t} F$};
  \node (7) [below of=5] { $C : \n{s} D E F$};

  \path[every node/.style={font=\sffamily\scriptsize}]
    (1) edge node [left] {$v(s)$} (4)
    (3) edge node [right] {$v(s)$} (4)
    (4) edge node [left] {$v(t)$} (7)
    (6) edge node [right] {$v(t)$} (7);
\end{tikzpicture}
}
\end{minipage}
& $\Rrightarrow$ &
\begin{minipage}{0.53\textwidth}
\scalebox{0.85}{
\begin{tikzpicture}[->,every node/.style={font=\sffamily\small}]
  \node (1) { $C'= C_2 : \n{s} E$};
\end{tikzpicture}
}
\end{minipage}\\
& & \\
\hline
\end{tabular}
\caption{Local transformation rules for resolution proof trees.}
\label{fig:rulestree}
\end{figure}
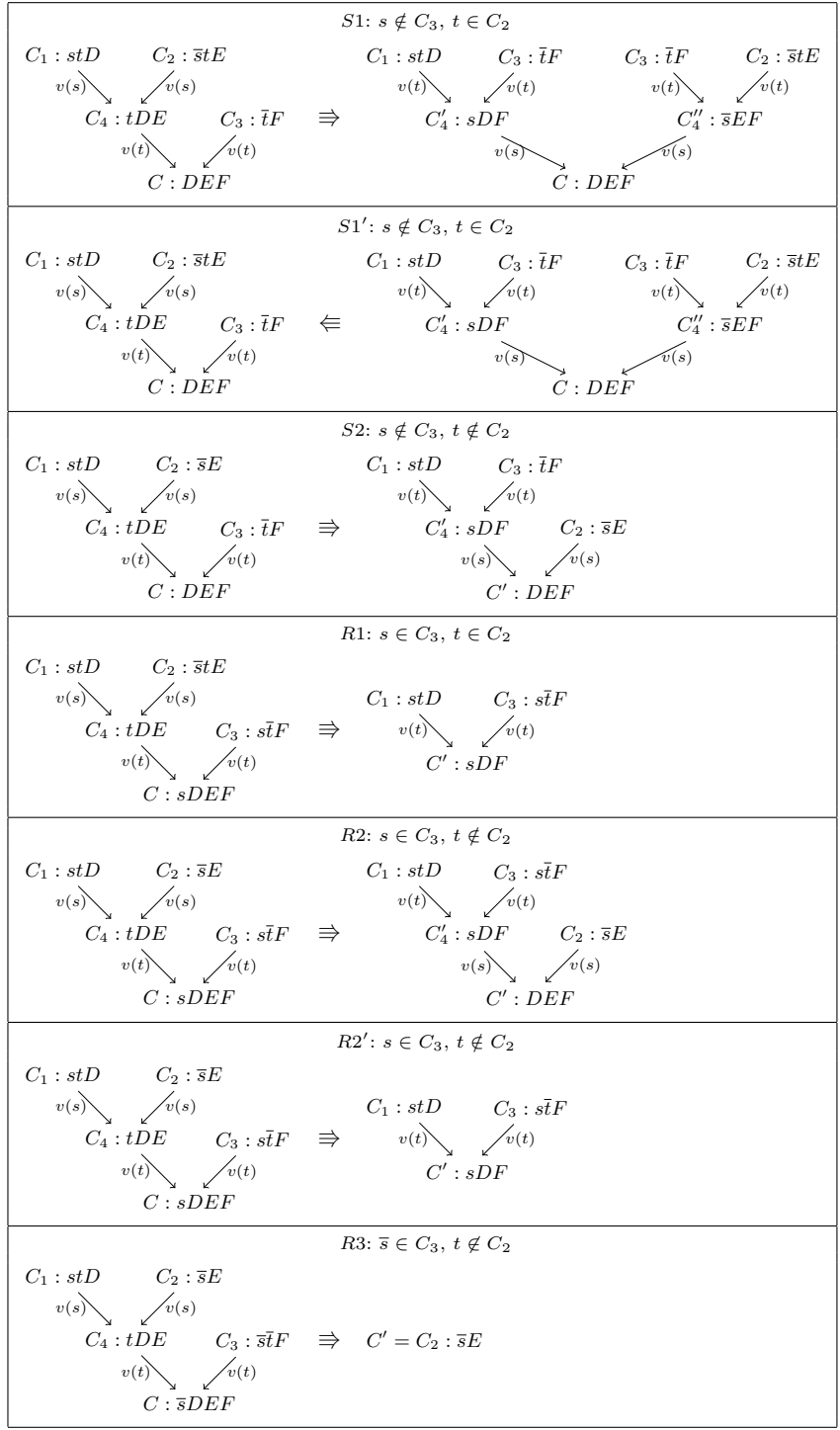

Fig.~\ref{fig:rulestree} shows a set of proof transformation rules. Each rule is 
associated with a unique context, and, conversely, each context can be
mapped to at least one rule (i.e., the set of rules is exhaustive, modulo
symmetry, for every possible context).
A first property that characterizes the set of rules is \emph{locality}: 
only the limited information represented by a context is in fact 
needed to determine which rule is applicable. A second property is \emph{strengthening}:
the rules either keep the context root unchanged or turn it into a logically stronger formula.

The classification of rules into $S$ (swapping) and $R$ (reducing) depends on 
the effect of the rules on the context rooted in $C$: $S1$ and $S2$ swap the two resolution 
steps in the context
without modifying $C$, 
while $R1$, $R2$,  $R2'$ and $R3$ replace $C$ with a new $C'$ such that 
$C' \subseteq C$; in other words, the $R$ rules generate
subproofs with stronger roots. 

The influence of the $S$ rules does not extend beyond the context where they are applied, while that of the $R$ rules
possibly propagates down to the global root.
The $R$ rules essentially simplify the proof and their effect cannot be undone, while an application of an $S$ rule can be reversed.
In particular, the effect of rule $S2$ can be canceled out simply by means of another application of
the same $S2$. $S1$ has $S1'$ as its inverse (notice the direction
of the arrow); $S1'$ is actually a derived rule, since it corresponds to the sequential application of $S2$ and $R2$. 

The rules $R2$ and $R2'$ are associated with the same context; they respectively behave as $S2$ (with an additional simplification of the root)
and $R1$. The decision whether to apply either rule depends on the overall goal of the transformation. 
Note that the application of rule $R2$ to a context turns it into a new context which matches rule $S1$.  
   
\subsection{Extension to Resolution Proof DAGs}
\label{subsec:ruledags}

If the  proof to be transformed is a DAG rather than a tree, some constraints are necessary on the application of the rules.

Consider rules $S1,S1',S2,R2$, and suppose clause $C_4$ is involved in more than one resolution step, having thus at least
another resolvent $C_5$ besides $C$. If $C_4$ is modified by a rule, 
it is not guaranteed that the correctness of the resolution step having $C_5$ as resolvent (and in turn of the resolution steps on the path from $C_5$ to the global root) is preserved. 
This problem does not concern clauses $C_1$, $C_2$, $C_3$ and the subproofs rooted in them, which are not changed by any rule.

A simple solution consists in creating a copy of $C_4$, to which all resolvents
of $C_4$ besides $C$ are assigned, so that $C_4$ is left with exactly one resolvent; at that point
any modification to $C_4$ will affect only the context rooted in $C$.
Since duplications increase the size of the proof, they should be carried out with moderation (see \S\ref{sec:heuri}). 

\begin{figure}[!htp]
\centering
\renewcommand{\arraystretch}{0.34}
\begin{tabular}{|p{0.41\textwidth} p{0.03\textwidth} p{0.45\textwidth}|}
\hline
& & \\
\multicolumn{3}{|c|}{\scriptsize{ $S1$: $s \notin C_3$, $t \in C_2$}} \\
& & \\
\begin{minipage}{.41\textwidth}
\scalebox{0.85}{
\begin{tikzpicture}[->,every node/.style={font=\sffamily\small}]
  \node (1) {$C_1 : s t D$};
  \node (2) [right of=1] {};
  \node (3) [right of=2] {$C_2: \n{s} t E$};
  \node (4) [below of=2] {$C_4 : t D E$};
  \node (5) [right of=4] {};
  \node (6) [right of=5] {$C_3: \n{t} F$};
  \node (7) [below of=5] { $C : D E F$};
  \node (8) [below of=1] {};
  \node (9) [below of=8]{};

  \path[every node/.style={font=\sffamily\scriptsize}]
    (1) edge node [left] {$v(s)$} (4)
    (3) edge node [right] {$v(s)$} (4)
    (4) edge node [left] {$v(t)$} (7)
    (6) edge node [right] {$v(t)$} (7);
  \path[every node/.style={font=\sffamily\scriptsize}, dotted]
    (4) edge node [] {} (9);
\end{tikzpicture}
}
\end{minipage}
& $\Rrightarrow$ &
\begin{minipage}{.45\textwidth}
\scalebox{0.85}{
\begin{tikzpicture}[->,every node/.style={font=\sffamily\small}]
  \node (1) {$C_1: s t D$};
  \node (2) [right of=1] {};
  \node (3) [right of=2] {$C_3: \n{t} F$};
  \node (6) [right of=3] {};
  \node (7) [right of=6] {$C_2: \n{s} t E$};
  \node (8) [below of=2] {$C_4': s D F$};
  \node (9) [below of=3] {};
  \node (10) [below of=6] {$C_4'': \n{s} E F$};
  \node (11) [below of=9] {$C: D E F$};
  \node (12) [below of=1] {};
  \node (13) [left of=12] {$C_4$};
  \node (14) [below of=12] {};

  \path[every node/.style={font=\sffamily\scriptsize}]
    (1) edge node [left] {$v(t)$} (8)
    (3) edge node [right=1.7mm] {$v(t)$} (8)
    (3) edge node [left] {} (10)
    (7) edge node [right] {$v(t)$} (10)
    (8) edge node [left] {$v(s)$} (11)
    (10) edge node [right] {$v(s)$} (11);
    \path[every node/.style={font=\sffamily\scriptsize}, dotted]
    (1) edge node [] {} (13)
    (7) edge node [] {} (13)
    (13) edge node [] {} (14);
\end{tikzpicture}
}
\end{minipage}\\
& & \\
\hline
& & \\
\multicolumn{3}{|c|}{\scriptsize{ $S1'$: $s \notin C_3$, $t \in C_2$}} \\
& & \\
\begin{minipage}{.41\textwidth}
\scalebox{0.85}{
\begin{tikzpicture}[->,every node/.style={font=\sffamily\small}]
  \node (1) {$C_1 : s t D$};
  \node (2) [right of=1] {};
  \node (3) [right of=2] {$C_2: \n{s} t E$};
  \node (4) [below of=2] {$C_4 : t D E$};
  \node (5) [right of=4] {};
  \node (6) [right of=5] {$C_3: \n{t} F$};
  \node (7) [below of=5] { $C : D E F$};
  \node (8) [left of=4] {};
  \node (9) [left of=8] {$C_4'$};
  \node (10) [left of=7] {};
  \node (11) [left of=10] {};

  \path[every node/.style={font=\sffamily\scriptsize}]
    (1) edge node [left] {$v(s)$} (4)
    (3) edge node [right] {$v(s)$} (4)
    (4) edge node [left] {$v(t)$} (7)
    (6) edge node [right] {$v(t)$} (7);
  \path[every node/.style={font=\sffamily\scriptsize}, dotted]
    (1) edge node [] {} (9)
    (6) edge node [] {} (9)
    (9) edge node [] {} (11);
\end{tikzpicture}
}
\end{minipage}
& $\Lleftarrow$ &
\begin{minipage}{.45\textwidth}
\scalebox{0.85}{
\begin{tikzpicture}[->,every node/.style={font=\sffamily\small}]
  \node (1) {$C_1: s t D$};
  \node (2) [right of=1] {};
  \node (3) [right of=2] {$C_3: \n{t} F$};
  \node (6) [right of=3] {};
  \node (7) [right of=6] {$C_2: \n{s} t E$};
  \node (8) [below of=2] {$C_4': s D F$};
  \node (9) [below of=3] {};
  \node (10) [below of=6] {$C_4'': \n{s} E F$};
  \node (11) [below of=9] {$C: D E F$};
  \node (12) [left of=11] {};
  \node (13) [left of=12] {};

  \path[every node/.style={font=\sffamily\scriptsize}]
    (1) edge node [left] {$v(t)$} (8)
    (3) edge node [right=1.7mm] {$v(t)$} (8)
    (3) edge node [left] {} (10)
    (7) edge node [right] {$v(t)$} (10)
    (8) edge node [left] {$v(s)$} (11)
    (10) edge node [right] {$v(s)$} (11);
  \path[every node/.style={font=\sffamily\scriptsize}, dotted]
    (8) edge node [right] {} (13);
\end{tikzpicture}
}
\end{minipage}\\
& & \\
\hline
& & \\
\multicolumn{3}{|c|}{\scriptsize{ $S2$: $s \notin C_3$, $t \notin C_2$}} \\
& & \\
\begin{minipage}{.41\textwidth}
\scalebox{0.85}{
\begin{tikzpicture}[->,every node/.style={font=\sffamily\small}]
  \node (1) {$C_1 : s t D$};
  \node (2) [right of=1] {};
  \node (3) [right of=2] {$C_2: \n{s} E$};
  \node (4) [below of=2] {$C_4 : t D E$};
  \node (5) [right of=4] {};
  \node (6) [right of=5] {$C_3: \n{t} F$};
  \node (7) [below of=5] { $C : D E F$};
      \node (8) [below of=1] {};
  \node (9) [below of=8]{};

  \path[every node/.style={font=\sffamily\scriptsize}]
    (1) edge node [left] {$v(s)$} (4)
    (3) edge node [right] {$v(s)$} (4)
    (4) edge node [left] {$v(t)$} (7)
    (6) edge node [right] {$v(t)$} (7);
   \path[every node/.style={font=\sffamily\scriptsize}, dotted]
    (4) edge node [] {} (9);  
\end{tikzpicture}
}
\end{minipage}
& $\Rrightarrow$ &
\begin{minipage}{.45\textwidth}
\scalebox{0.85}{
\begin{tikzpicture}[->,every node/.style={font=\sffamily\small}]
  \node (1) {$C_1 : s t D$};
  \node (2) [right of=1] {};
  \node (3) [right of=2] {$C_3: \n{t} F$};
  \node (4) [below of=2] {$C_4': s D F$};
  \node (5) [right of=4] {};
  \node (6) [right of=5] {$C_2: \n{s} E$};
  \node (7) [below of=5] { $C' : D E F$};
    \node (12) [below of=1] {};
  \node (13) [left of=12] {$C_4$};
  \node (14) [below of=12] {};

  \path[every node/.style={font=\sffamily\scriptsize}]
    (1) edge node [left] {$v(t)$} (4)
    (3) edge node [right] {$v(t)$} (4)
    (4) edge node [left] {$v(s)$} (7)
    (6) edge node [right] {$v(s)$} (7);
      \path[every node/.style={font=\sffamily\scriptsize}, dotted]
          (1) edge node [] {} (13)
    (6) edge node [] {} (13)
    (13) edge node [] {} (14);
\end{tikzpicture}
}
\end{minipage}\\
& & \\
\hline
& & \\
\multicolumn{3}{|c|}{\scriptsize{ $R1$: $s \in C_3$, $t \in C_2$}} \\
& & \\
\begin{minipage}{.41\textwidth}
\scalebox{0.85}{
\begin{tikzpicture}[->,every node/.style={font=\sffamily\small}]
  \node (1) {$C_1 : s t D$};
  \node (2) [right of=1] {};
  \node (3) [right of=2] {$C_2: \n{s} t E$};
  \node (4) [below of=2] {$C_4 : t D E$};
  \node (5) [right of=4] {};
  \node (6) [right of=5] {$C_3: s \n{t} F$};
  \node (7) [below of=5] { $C : s D E F$};
    \node (8) [below of=1] {};
  \node (9) [below of=8]{};

  \path[every node/.style={font=\sffamily\scriptsize}]
    (1) edge node [left] {$v(s)$} (4)
    (3) edge node [right] {$v(s)$} (4)
    (4) edge node [left] {$v(t)$} (7)
    (6) edge node [right] {$v(t)$} (7);
      \path[every node/.style={font=\sffamily\scriptsize}, dotted]
    (4) edge node [] {} (9);  
\end{tikzpicture}
}
\end{minipage}
& $\Rrightarrow$ &
\begin{minipage}{.45\textwidth}
\scalebox{0.85}{
\begin{tikzpicture}[->,every node/.style={font=\sffamily\small}]
  \node (1) {$C_2: \n{s} t E$};
  \node (2) [right of=1] {};
  \node (3) [right of=2] {$C_1: s t D$};
  \node (4) [right of=3] {};
  \node (5) [right of=4] {$C_3: s \n{t} F$ };
  \node (6) [below of=2] {$C_4$};
  \node (7) [below of=4] {$C': s D F$};
      \node (8) [below of=1] {};
  \node (9) [below of=8]{};

  \path[every node/.style={font=\sffamily\scriptsize}]
    (3) edge node [left] {$v(t)$} (7)
    (5) edge node [right] {$v(t)$} (7);
  \path[every node/.style={font=\sffamily\scriptsize}, dotted]
    (1) edge node [left] {$v(s)$} (6)
    (3) edge node [below right] {$v(s)$} (6)
        (6) edge node [] {} (9); 
\end{tikzpicture}
}
\end{minipage} \\
& & \\
\hline
& & \\
\multicolumn{3}{|c|}{\scriptsize{ $R2$: $s \in C_3$, $t \notin C_2$}} \\
& & \\
\begin{minipage}{.41\textwidth}
\scalebox{0.85}{
\begin{tikzpicture}[->,every node/.style={font=\sffamily\small}]
  \node (1) {$C_1 : s t D$};
  \node (2) [right of=1] {};
  \node (3) [right of=2] {$C_2: \n{s} E$};
  \node (4) [below of=2] {$C_4 : t D E$};
  \node (5) [right of=4] {};
  \node (6) [right of=5] {$C_3: s \n{t} F$};
  \node (7) [below of=5] { $C : s D E F$};
      \node (8) [below of=1] {};
  \node (9) [below of=8]{};

  \path[every node/.style={font=\sffamily\scriptsize}]
    (1) edge node [left] {$v(s)$} (4)
    (3) edge node [right] {$v(s)$} (4)
    (4) edge node [left] {$v(t)$} (7)
    (6) edge node [right] {$v(t)$} (7);
   \path[every node/.style={font=\sffamily\scriptsize}, dotted]
    (4) edge node [] {} (9);  
\end{tikzpicture}
}
\end{minipage}
& $\Rrightarrow$ &
\begin{minipage}{.45\textwidth}
\scalebox{0.85}{
\begin{tikzpicture}[->,every node/.style={font=\sffamily\small}]
  \node (1) {$C_1 : s t D$};
  \node (2) [right of=1] {};
  \node (3) [right of=2] {$C_3: s \n{t} F$};
  \node (4) [below of=2] {$C_4': s D F$};
  \node (5) [right of=4] {};
  \node (6) [right of=5] {$C_2: \n{s} E$};
  \node (7) [below of=5] { $C' : D E F$};
    \node (12) [below of=1] {};
  \node (13) [left of=12] {$C_4$};
  \node (14) [below of=12] {};

  \path[every node/.style={font=\sffamily\scriptsize}]
    (1) edge node [left] {$v(t)$} (4)
    (3) edge node [right] {$v(t)$} (4)
    (4) edge node [left] {$v(s)$} (7)
    (6) edge node [right] {$v(s)$} (7);
      \path[every node/.style={font=\sffamily\scriptsize}, dotted]
          (1) edge node [] {} (13)
    (6) edge node [] {} (13)
    (13) edge node [] {} (14);
\end{tikzpicture}
}
\end{minipage}\\
& & \\
\hline
& & \\
\multicolumn{3}{|c|}{\scriptsize{ $R2'$: $s \in C_3$, $t \notin C_2$}} \\
& & \\
\begin{minipage}{.41\textwidth}
\scalebox{0.85}{
\begin{tikzpicture}[->,every node/.style={font=\sffamily\small}]
  \node (1) {$C_1 : s t D$};
  \node (2) [right of=1] {};
  \node (3) [right of=2] {$C_2: \n{s} E$};
  \node (4) [below of=2] {$C_4 : t D E$};
  \node (5) [right of=4] {};
  \node (6) [right of=5] {$C_3: s \n{t} F$};
  \node (7) [below of=5] { $C : s D E F$};
      \node (8) [below of=1] {};
  \node (9) [below of=8]{};

  \path[every node/.style={font=\sffamily\scriptsize}]
    (1) edge node [left] {$v(s)$} (4)
    (3) edge node [right] {$v(s)$} (4)
    (4) edge node [left] {$v(t)$} (7)
    (6) edge node [right] {$v(t)$} (7);
   \path[every node/.style={font=\sffamily\scriptsize}, dotted]
    (4) edge node [] {} (9);  
\end{tikzpicture}
}
\end{minipage}
& $\Rrightarrow$ &
\begin{minipage}{.45\textwidth}
\scalebox{0.85}{
\begin{tikzpicture}[->,every node/.style={font=\sffamily\small}]
  \node (1) {$C_2: \n{s} E$};
  \node (2) [right of=1] {};
  \node (3) [right of=2] {$C_1: s t D$};
  \node (4) [right of=3] {};
  \node (5) [right of=4] {$C_3: s \n{t} F$ };
  \node (6) [below of=2] {$C_4$};
  \node (7) [below of=4] {$C': s D F$};
      \node (8) [below of=1] {};
  \node (9) [below of=8]{};

  \path[every node/.style={font=\sffamily\scriptsize}]
    (3) edge node [left] {$v(t)$} (7)
    (5) edge node [right] {$v(t)$} (7);
  \path[every node/.style={font=\sffamily\scriptsize}, dotted]
    (1) edge node [left] {$v(s)$} (6)
    (3) edge node [below right] {$v(s)$} (6)
        (6) edge node [] {} (9); 
\end{tikzpicture}
}
\end{minipage} \\
& & \\
\hline
& & \\
\multicolumn{3}{|c|}{\scriptsize{ $R3$: $\n{s} \in C_3$, $t \not\in C_2$}} \\
& & \\
\begin{minipage}{.41\textwidth}
\scalebox{0.85}{
\begin{tikzpicture}[->,every node/.style={font=\sffamily\small}]
  \node (1) {$C_1 : s t D$};
  \node (2) [right of=1] {};
  \node (3) [right of=2] {$C_2: \n{s} E$};
  \node (4) [below of=2] {$C_4 : t D E$};
  \node (5) [right of=4] {};
  \node (6) [right of=5] {$C_3: \n{s} \n{t} F$};
  \node (7) [below of=5] { $C : \n{s} D E F$};
     \node (8) [below of=1] {};
  \node (9) [below of=8]{};

  \path[every node/.style={font=\sffamily\scriptsize}]
    (1) edge node [left] {$v(s)$} (4)
    (3) edge node [right] {$v(s)$} (4)
    (4) edge node [left] {$v(t)$} (7)
    (6) edge node [right] {$v(t)$} (7);
   \path[every node/.style={font=\sffamily\scriptsize}, dotted]
    (4) edge node [] {} (9);  
\end{tikzpicture}
}
\end{minipage}
& $\Rrightarrow$ &
\begin{minipage}{.45\textwidth}
\scalebox{0.85}{
\begin{tikzpicture}[->,every node/.style={font=\sffamily\small}]
  \node (1) {$C_1: s t D$};
  \node (2) [right of=1] {};
  \node (3) [right of=2] {$C' = C_2: \n{s} E$};
  \node (6) [below of=2] {$C_4$};
      \node (8) [below of=1] {};
  \node (9) [below of=8]{};

  \path[every node/.style={font=\sffamily\scriptsize}, dotted]
    (1) edge node [left] {$v(s)$} (6)
    (3) edge node [right] {$v(s)$} (6)
        (6) edge node [] {} (9); 
\end{tikzpicture}
}
\end{minipage} \\
& & \\
\hline
\end{tabular}
\caption{Local transformation rules for resolution proof DAGs.}
\label{fig:rulesdag}
\end{figure}
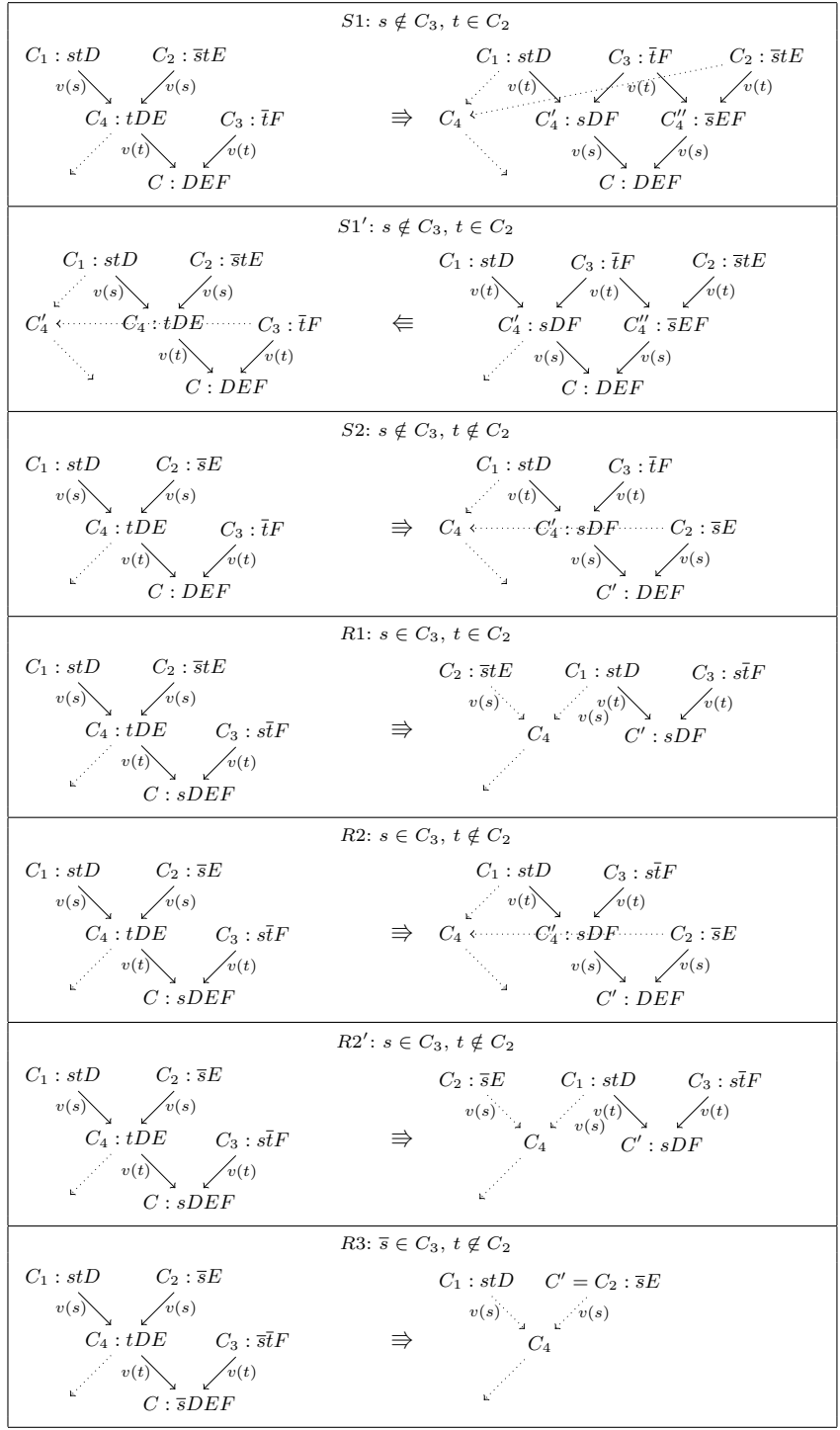

A more efficient alternative exists in case of rules $R1$, $R2'$, $R3$, where $C_4$ is detached by the context rooted in $C$ and loses $C$ as resolvent, but
keeps the other resolvents (if any). The effect of the transformation rules is shown in Fig.~\ref{fig:rulesdag}: the presence of additional resolvents for $C_4$
is denoted by a dotted arrow.

\subsection{Soundness of the Local Transformation Framework}
\label{subsec:sound}

In this section we first prove that the rewriting rules preserve the legality of the subproofs rooted in the contexts 
where the rules are applied; then we discuss how the rules affect the global proof and what steps must be taken to maintain it legal. 

\paragraph{Effect on a context.}
Based on the following observations, we claim
that after a single application of a rule to a context with root $C$, the 
legal subproof rooted in $C$ is replaced by a legal subproof rooted 
in $C' \subseteq C$.

Refer to Fig.~\ref{fig:rulesdag}. No additional subproofs are introduced by the rules and no modifications are brought to the subproofs rooted 
in $C_1, C_2, C_3$, which are simply recombined or detached from the context.
As for the $S$ rules, $C_4$ is either replaced by the resolvent of $C_1,C_3$ ($S2$) or by the resolvent of the resolvents of $C_1,C_3$ and $C_3,C_2$ 
($S1$, where a new clause $C_4''=Res_{v(s)}(C_2,C_3)$ is also introduced). Note that in both cases $C$ is not modified.
The $R$ rules instead yield a more substantial change in the form of a stronger context root $C' \subseteq C$:
\begin{itemize}
\item In $R1$ and $R2'$, the subproofs with root $C_1$ and $C_3$ are combined to obtain a subproof with root $s D F \subseteq s D E F$.
\item $R2$ has a swap effect similar to $S2$, but replaces the root $s D E F$ with $D E F$, removing a single literal.
\item In $R3$, the whole subproof is substituted by the subproof rooted in $C_2=\n{s} E$, which subsumes  $C=\n{s} D E F$. 
\end{itemize}
All the above transformations involve updating the relevant clauses by means of sound applications of the resolution rule.

\paragraph{Effect on the global proof.}
The application of a rule to a context yields a legal subproof rooted in a clause $C' \subseteq C$;
however, the global proof could turn into an illegal one.
In fact, the deletion of literals from $C$ affects the sequence of resolution steps that extends from  $C$ to the global root: 
some of these steps might become superfluous, because they resolve upon a variable which was introduced by $C$ (but does not appear in $C'$), 
and they should be appropriately removed. In the same way, the elimination of a resolution step could itself lead to the disappearance of more literal occurrences, 
leading to a chain reaction. 

The following Alg.~\ref{alg:prop}, \emph{SubsumptionPropagation}, has the purpose of propagating the effect 
of the replacement of $C$ by $C' \subseteq C$ along the path leading 
from $C$ to the global root.

The algorithm restructures the proof in a top-down manner 
analyzing the sequence of resolution steps to ensure their correctness 
while propagating the effect of the initial subsumption. We  
prove that, after an execution of SubsumptionPropagation following the application of an $R$ rule to a legal proof, 
the result is still a legal proof.

\begin{algorithm}
\SetAlgoNoLine
\LinesNumbered
\DontPrintSemicolon
\KwIn{A legal proof modified by an $R$ rule}
\KwOut{A legal proof}
\KwData{$W$: set of nodes reachable from $C'$, $V$: set of visited nodes}
\Begin{
$V \leftarrow \emptyset$\;
Determine $W$, e.g. through a visit from $C'$\;
\While{$W \setminus V \neq \emptyset$}{
Choose $n \in W \setminus V$ such that: \\ 
$(n^+ \in W$ \text{or} $n^+ \notin W)$ \text{and} $(n^- \in W$ \text{or} $n^- \notin W)$\;
$V \leftarrow V \cup \{n\}$\;
$p\leftarrow piv(n)$\;
\uIf{$p \in C(n^+)$ {and} $\n{p} \in C(n^-)$}{
$C(n) \leftarrow Res_p(C(n^+),C(n^-))$
}
\uElseIf{$p \notin C(n^+)$ {and} $\n{p} \in C(n^-)$}{
Substitute $n$ with $n^+$\;
}
\uElseIf{$p \in C(n^+)$ {and} $\n{p} \notin C(n^-)$}{
Substitute $n$ with $n^-$\;
}
\uElseIf{$p \notin C(n^+)$ {and} $\n{p} \notin C(n^-)$}{
Heuristically choose a parent, replace $n$ with it\;
}
}
}
\caption{SubsumptionPropagation.}
\label{alg:prop}
\end{algorithm}
The idea at the base of the algorithm reflects the mechanisms of the restructuring procedures first
proposed in~\cite{BFHSS08,DKPW10}:
\begin{enumerate}
\item It determines the effect range of the 
      substitution of $C$ by $C'$, which corresponds 
      to the set of nodes reachable from the node labeled by $C'$.
\item It analyzes, one by one, all reachable nodes; 
      it is necessary that the antecedents of a node $n$ 
      have already been visited (and possibly modified), 
      in order to guarantee a correct propagation 
      of the modifications to $n$.
\item Due to the potential vanishing of literals from clauses, 
      it might happen that in some resolution step the pivot 
      is not present in both antecedents anymore; if that 
      is the case, the resolution step is deleted, by replacing 
      the resolvent with the antecedent devoid of the pivot 
      (if the pivot is missing in both antecedents, 
       either of them is arbitrarily chosen), otherwise,
       the resolution step is kept and the resolvent clause updated. 
      At the graph level, $n$ is substituted by $n^+$ or $n^-$, 
      assigning the children of $n$ (if any) to it.
\end{enumerate}  
\begin{figure}[!ht]
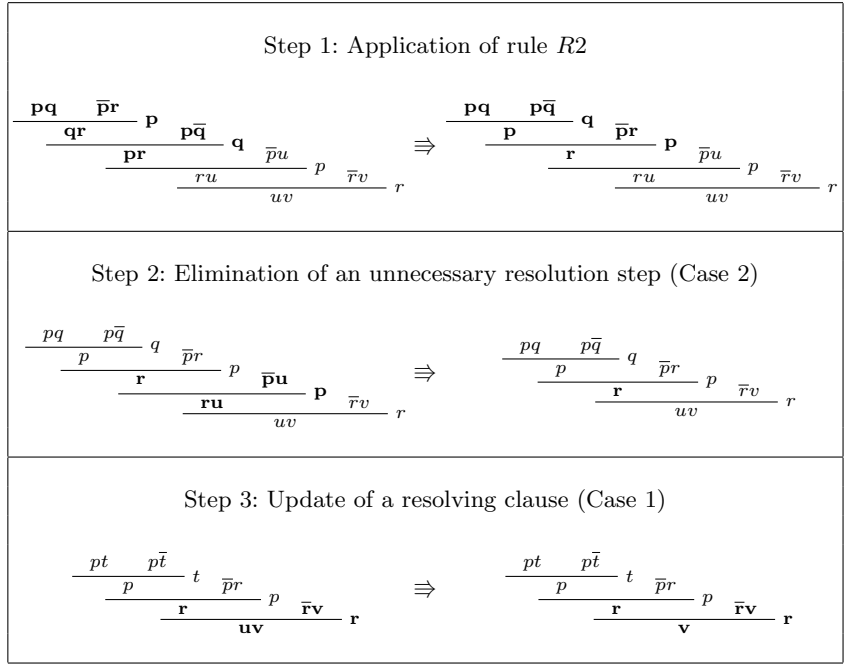

\centering
\begin{tabular}{|lcr|}
\hline
& & \\
\multicolumn{3}{|c|}{Step 1: Application of rule $R2$} \\
& & \\
\begin{minipage}{.43\textwidth}
\scriptsize
\def\defaultHypSeparation{\hskip .02in}
\begin{prooftree}
\AxiomC{ $\mathbf{p} \mathbf{q}$ }
\AxiomC{ $\mathbf{\n{p}} \mathbf{r}$ }
\RightLabel{$\mathbf{p}$}
\BinaryInfC{ $\mathbf{q} \mathbf{r}$ }
\AxiomC{ $\mathbf{p} \mathbf{\n{q}} $ }
\RightLabel{$\mathbf{q}$}
\BinaryInfC{ $\mathbf{p} \mathbf{r} $ }
\AxiomC{ $\n{p} u$ }
\RightLabel{$p$}
\BinaryInfC{ $r u$ }
\AxiomC{ $\n{r} v$ }
\RightLabel{$r$}
\BinaryInfC{ $u v$ }
\end{prooftree}
\end{minipage}
& $\Rrightarrow$ &
\begin{minipage}{.43\textwidth}
\scriptsize
\def\defaultHypSeparation{\hskip .02in}
\begin{prooftree}
\AxiomC{ $\mathbf{p} \mathbf{q}$ }
\AxiomC{ $\mathbf{p} \mathbf{\n{q}} $ }
\RightLabel{$\mathbf{q}$}
\BinaryInfC{ $\mathbf{p}$ }
\AxiomC{ $\mathbf{\n{p}} \mathbf{r} $ }
\RightLabel{$\mathbf{p}$}
\BinaryInfC{ $ \mathbf{r} $ }
\AxiomC{ $\n{p} u$ }
\RightLabel{$p$}
\BinaryInfC{ $r u$ }
\AxiomC{ $\n{r} v$ }
\RightLabel{$r$}
\BinaryInfC{ $u v$ }
\end{prooftree}
\end{minipage} \\
& & \\
\hline
& & \\
\multicolumn{3}{|c|}{Step 2: Elimination of an unnecessary resolution step (Case 2)} \\
& & \\
\begin{minipage}{.43\textwidth}
\scriptsize
\def\defaultHypSeparation{\hskip .02in}
\begin{prooftree}
\AxiomC{ $p q$ }
\AxiomC{ $p \n{q} $ }
\RightLabel{$q$}
\BinaryInfC{ $p$ }
\AxiomC{ $\n{p} r $ }
\RightLabel{$p$}
\BinaryInfC{ $ \mathbf{r} $ }
\AxiomC{ $\mathbf{\n{p}} \mathbf{u}$ }
\RightLabel{$\mathbf{p}$}
\BinaryInfC{ $\mathbf{r} \mathbf{u}$ }
\AxiomC{ $\n{r} v$ }
\RightLabel{$r$}
\BinaryInfC{ $u v$ }
\end{prooftree}
\end{minipage}
& $\Rrightarrow$ &
\begin{minipage}{.43\textwidth}
\scriptsize
\def\defaultHypSeparation{\hskip .02in}
\begin{prooftree}
\AxiomC{ $p q$ }
\AxiomC{ $p \n{q} $ }
\RightLabel{$q$}
\BinaryInfC{ $p$ }
\AxiomC{ $\n{p} r $ }
\RightLabel{$p$}
\BinaryInfC{ $ \mathbf{r} $ }
\AxiomC{ $\n{r} v$ }
\RightLabel{$r$}
\BinaryInfC{ $u v$ }
\end{prooftree}
\end{minipage} \\
& & \\
\hline
& & \\
\multicolumn{3}{|c|}{Step 3: Update of a resolving clause (Case 1)} \\
& & \\
\begin{minipage}{.43\textwidth}
\scriptsize
\def\defaultHypSeparation{\hskip .02in}
\begin{prooftree}
\AxiomC{ $p t$ }
\AxiomC{ $p \n{t} $ }
\RightLabel{$t$}
\BinaryInfC{ $p$ }
\AxiomC{ $\n{p} r $ }
\RightLabel{$p$}
\BinaryInfC{ $ \mathbf{r} $ }
\AxiomC{ $\mathbf{\n{r}} \mathbf{v}$ }
\RightLabel{$\mathbf{r}$}
\BinaryInfC{ $\mathbf{u} \mathbf{v}$ }
\end{prooftree}
\end{minipage}
& $\Rrightarrow$ &
\begin{minipage}{.43\textwidth}
\scriptsize
\def\defaultHypSeparation{\hskip .02in}
\begin{prooftree}
\AxiomC{ $p t$ }
\AxiomC{ $p \n{t} $ }
\RightLabel{$t$}
\BinaryInfC{ $p$ }
\AxiomC{ $\n{p} r $ }
\RightLabel{$p$}
\BinaryInfC{ $ \mathbf{r} $ }
\AxiomC{ $\mathbf{\n{r}} \mathbf{v}$ }
\RightLabel{$\mathbf{r}$}
\BinaryInfC{ $\mathbf{v}$ }
\end{prooftree}
\end{minipage} \\
& & \\
\hline
\end{tabular}
\caption{Example of rule application and subsumption propagation.}
\label{fig:subsex}
\end{figure}
\begin{thm}
\label{thm:subprop}
Assume a legal proof $P$.  The application of an  $R$ rule, followed by an execution of  SubsumptionPropagation, yields a legal proof $P'$,
whose new global root subsumes the previous one.
\begin{proof}[by structural induction]
\paragraph{\textbf{Base case}.} Assume an $R$ rule is applied to a  context rooted in a clause $C$; $C$ is replaced by $C' \subseteq C$ and 
the subproof rooted in $C'$ is legal, as previously shown. The subproofs rooted in the clauses of nodes not reachable from $C$ are not affected and thus remain legal.
\paragraph{\textbf{Inductive step}.} 
All nodes reachable from $C$ are visited; in particular, a node $n$ is visited after its reachable parents. 
By inductive hypothesis $C'(n^{+}) \subseteq C(n^+)$, $C'(n^-)\subseteq C(n^-)$ and
the subproofs rooted in $C'(n^+)$ and $C'(n^-)$ are legal.
We show that, after visiting $n$, $C'(n) \subseteq C(n)$ and the subproof rooted in $C'(n)$ is legal.
Let $p = piv(n)$.
We have three possibilities:
\begin{itemize}
\item Case 1: the pivot still appears both in $C'(n^+)$ and in $C'(n^-)$; $C'(n)=$ \lb $Res_{p}(C'(n^+),C'(n^-))$, thus $C'(n)\subseteq C(n)$.
\item Case 2: the pivot is present only in one antecedent, let us say $C'(n^+)$; the subproof rooted in $C(n)$ is replaced by the one rooted in $C'(n^-)$ (legal by hypothesis). 
But $C'(n)=C'(n^-) \subseteq C(n)$ since $C'(n^-)$ does not contain the pivot.
\item Case 3: the pivot is not present in either antecedent. Same reasoning as for Case 2, but arbitrarily choosing an antecedent for the substitution.
\end{itemize}
In all three cases the subproof rooted in $C'(n)$ is legal and $C'(n)\subseteq C(n)$.
\end{proof}
\end{thm}
Fig.~\ref{fig:subsex} shows the effect of $R2$ and the subsequent application of SubsumptionPropagation on a small proof.

\subsection{A Transformation Meta-Algorithm}
\label{subsec:meta}

The Local Transformation Framework defined by our rules leaves to the user the
flexibility of choosing a particular {\em strategy}
and a {\em termination criterion} for their application.

Whenever a sizeable amount of rules has to be applied, rather than running SubsumptionPropagation multiple times, 
it is more efficient to combine
the application of all rules and the propagation of the modifications into a single
traversal of the proof.

Alg.~\ref{alg:rectraloop}, \emph{TransformAndReconstruct}, illustrates this approach. At first it performs a topological sorting of the proof (line 2), in order to ensure that each 
node is visited after its parents. Then it analyzes one node at a time, checking if the corresponding resolution 
step is still sound (line 6). If the resolution step is sound, it updates the resolvent clause, determining the node contexts (if any) and the 
associated rules. At most one rule is applied, and the decision is based on local heuristic considerations (line 9).
If the resolution step is not sound and either antecedent does not contain the pivot (lines 11, 13, 15), then the resolution step is removed  by 
replacing the resolvent with that antecedent (which, missing the pivot, subsumes the resolvent); at the graph level, $n$ is substituted by $n^+$ or $n^-$.

Note that the antecedent not responsible for the substitution might have lost all its resolvents and thus does not contribute 
to the proof anymore; in that case it is pruned away, together with the portion of the subproof rooted in it which has become 
detached from the global proof. 

A key point of the algorithm is the call to \emph{ApplyRule(left context, right context)}: 
this method heuristically chooses at most one context (possibly none) rooted in $n$ and applies the corresponding rule.
The instantiation of ApplyRule with different procedures yields concrete algorithms suitable for particular applications,
as illustrated in the next sections.

Based on the above observations and on Theorem~\ref{thm:subprop}, we have the following result:
\begin{thm}
TransformAndReconstruct outputs a legal proof.
\end{thm}
\begin{algorithm}[h!] 
\SetAlgoNoLine
\LinesNumbered
\DontPrintSemicolon
\KwIn{A legal proof, an instance of \emph{ApplyRule}}
\KwOut{A legal proof}
\KwData{$TS$: nodes topological sorting vector}
\Begin{
  $TS \leftarrow$ topological\_{}sorting\_top\_down(proof)\;
  \ForEach{$n \in TS$}{
    \uIf{$n$ {is not a leaf}}{
    $p\leftarrow piv(n)$\;
      \uIf{$\n{p} \in C(n^-)$ {and} $p \in C(n^+)$}{
	$C(n) \leftarrow Res_{p}(C(n^-),C(n^+))$\;
	Determine left context $lc$ of $n$, if any\;
	Determine right context $rc$ of $n$, if any\;
	ApplyRule($rc,lc$)\;
      }
      \uElseIf{$\n{p} \notin C(n^-)$ {and} $p \in C(n^+)$}{
	Substitute $n$ with $n^-$\;
      }
      \uElseIf{$\n{p} \in C(n^-)$ {and} $p \notin C(n^+)$}{
	Substitute $n$ with $n^+$\;
      }
      \uElseIf{$\n{p} \notin C(n^-)$ {and} $p \notin C(n^+)$}{
	Heuristically choose a parent, substitute $n$ with it\;
      }
    }
  }
}
\caption{TransformAndReconstruct.}
\label{alg:rectraloop}
\end{algorithm}

\section{Proof Compression}
\label{sec:compr}

Resolution proofs, as generated by modern solvers, 
find application in many verification techniques.
In most cases, the size of the proofs affects the efficiency of the methods 
in which they are used. It is known that the size of a resolution
proof can grow exponentially with respect to the size of the input formula: even 
when proofs are representable in a manageable memory space, it might be crucial for 
efficiency to reduce or compress them as much as possible. Several compression 
technique have been developed and can be found in literature, ranging from 
memoization of common subproofs to partial regularization \cite{A06,A08,S07,C10,BFHSS08,DKPW10,FMP11}; 
however, since the problem of finding a minimum proof is NP-hard, it is still 
an open challenge to design heuristics capable of obtaining good reduction in practical situations. 

This section discusses algorithms aimed at compressing proofs.
We identify two kinds of \emph{redundancies} in resolution proofs
and present a set of post-processing techniques aimed at removing them;
the techniques are independent from the way the refutation is produced and
can be applied to an arbitrary resolution proof of unsatisfiability. 
We also illustrate how to combine these algorithms in an effective manner, and show the results of experimenting
on a collection of SAT and SMT benchmarks.

We do not address directly the problem of core minimization, 
that is nonetheless achieved as a side effect of proof reduction.
A rich literature exists on techniques aimed at obtaining a minimum 
(a $\Sigma_2$-complete problem), minimal ($D^{P}$-complete), or 
small unsatisfiable core, that is a subset of the initial set of 
clauses that is still unsatisfiable~\cite{LMS04,CGS07,ZM03,OMASM04,H05,B03,DHN06,GMP07,MLAMS05}. 

\subsection{Proof Redundancies}
\label{subsec:red}
This paper focuses  on  two particular kinds of redundancies in resolution proofs.

The first one stems from the observation that, 
along each path from a leaf to the root, it is unnecessary to resolve upon a certain 
pivot more than once. The proof can be simplified, for example by keeping 
(for a given variable and a path) only the resolution step closest to the root, while 
cutting the others away. In the literature, a proof such that each variable is used as a pivot 
at most once along each path from a leaf to the root is said to be {\em regular}~\cite{Tse68}.

The second kind of redundancy is related to the content of a proof. It might be the case that
there exist multiple nodes associated with equal clauses; such nodes can be merged, keeping only
one pair of parents and grouping together all the children. In particular, 
 we call a proof \emph{compact} if $C(n_i) = C(n_j) \implies i = j$ for any $i,j$,  
 that is, different nodes are labeled by different clauses.
 
 \subsection{Proof Regularity}

In this section we discuss how to make a proof (partially) regular. 
We show how to employ Alg.~\ref{alg:rectraloop} for this purpose
and present two algorithms explicitly devised for regularization,
namely RecyclePivots~\cite{BFHSS08} and its refinement RecyclePivotsWithIntersection~\cite{FMP11}. 
We illustrate them individually and explain how they can be combined to obtain more powerful algorithms.

\subsubsection{Regularization in the Local Transformation Framework}
\label{sec:traalgo}

The  $R$ rules are, as a matter of fact, a means to perform a ``local'' regularization;
they are applied to contexts where a resolution step on a pivot $v(s)$ is immediately
followed by a reintroduction of the pivot with positive ($R1,R2,R'2$) or negative ($R3$) polarity
(see Fig.~\ref{fig:rulesdag}).

Resolving on $v(s)$ is redundant, since the newly introduced occurrence of the pivot will be later resolved upon
along the path to the global root; the $R$ rules have the effect of simplifying the context, possibly
pruning subproofs which do not contribute anymore to the global proof. Moreover, the rules replace the root of a context
with a stronger one, which allows to achieve further compression as shown below.

Consider, for example,  the following proof:

\medskip
\begin{tabular}{cr}
\begin{minipage}{.8\textwidth}
\begin{prooftree}
\scriptsize
\AxiomC{ $\mathbf{pq}$ }
\AxiomC{ $\mathbf{\n{p}o}$ }
\RightLabel{$p$}
\BinaryInfC{ $\mathbf{qo}$ }
\AxiomC{ $\mathbf{p\n{q}}$ }
\RightLabel{$q$}
\BinaryInfC{ $\mathbf{po}$ }
\AxiomC{ $qr$ }
\AxiomC{ $\n{p}\n{q}$ }
\RightLabel{$q$}
\BinaryInfC{ $\n{p}r$ }
\RightLabel{$p$}
\BinaryInfC{$or$}
\AxiomC{ $\n{o}s$ }
\RightLabel{$o$}
\BinaryInfC{$rs$}
\end{prooftree}
\end{minipage}
&
(1)
\end{tabular}
\medskip \\

\nin{}The highlighted context can be reduced via an application of $R2$ as follows:

\medskip
\begin{tabular}{cr}
\begin{minipage}{.8\textwidth}
\begin{prooftree}
\scriptsize
\AxiomC{ $\mathbf{pq}$ }
\AxiomC{ $\mathbf{p\n{q}}$ }
\RightLabel{$q$}
\BinaryInfC{ $\mathbf{p}$ }
\AxiomC{ $qr$ }
\AxiomC{ $\n{p}\n{q}$ }
\RightLabel{$q$}
\BinaryInfC{ $\n{p}r$ }
\RightLabel{$p$}
\BinaryInfC{$or$}
\AxiomC{ $\n{o}s$ }
\RightLabel{$o$}
\BinaryInfC{$rs$}
\end{prooftree}
\end{minipage}
&
(2)
\end{tabular}
\medskip \\

\nin{}The proof has become illegal as the literal $o$ is now not
introduced by any clause.
Since a stronger conclusion ($p \subset po$) has been derived, 
$o$ is now redundant and it can be eliminated all the way down to the
global root or up to the point it is reintroduced by some other resolution step.
In this example  $o$ can be safely removed together with the last
resolution step which also becomes redundant. The resulting legal 
(and stronger) proof becomes:

\medskip
\begin{tabular}{cr}
\begin{minipage}{.8\textwidth}
\begin{prooftree}
\scriptsize
\AxiomC{ $pq$ }
\AxiomC{ $p\n{q}$ }
\RightLabel{$q$}
\BinaryInfC{ $p$ }
\AxiomC{ $qr$ }
\AxiomC{ $\n{p}\n{q}$ }
\RightLabel{$q$}
\BinaryInfC{ $\n{p}r$ }
\RightLabel{$p$}
\BinaryInfC{$r$}
\end{prooftree}
\end{minipage}
&
(3)
\end{tabular}
\medskip \\

\nin{}At this stage no other $R$ rule can be directly applied to the proof.

Rule $S2$ does not perform any simplification on its own, however it is still used in our framework. 
Its contribution is to produce a ``shuffling'' effect in the proof, in order to create more 
chances for the $R$ rules to be applied.

Consider again our running example. $S2$ can be applied as follows:

\medskip
\begin{tabular}{cr}
\begin{minipage}{.8\textwidth}
\begin{prooftree}
\scriptsize
\AxiomC{ $pq$ }
\AxiomC{ $p\n{q}$ }
\RightLabel{$q$}
\BinaryInfC{ $\mathbf{p}$ }
\AxiomC{ $\mathbf{qr}$ }
\AxiomC{ $\mathbf{\n{p}\n{q}}$ }
\RightLabel{$q$}
\BinaryInfC{ $\mathbf{\n{p}r}$ }
\RightLabel{$p$}
\BinaryInfC{$\mathbf{r}$}
\end{prooftree}
\end{minipage}
&
(4)
\end{tabular}
\medskip \\

\medskip
\begin{tabular}{cr}
\begin{minipage}{.8\textwidth}
\scriptsize
\begin{prooftree}
\AxiomC{ $\mathbf{qr}$ }
\AxiomC{ $p\n{q}$ }
\AxiomC{ $pq$ }
\RightLabel{$q$}
\BinaryInfC{ $\mathbf{p}$ }
\AxiomC{ $\mathbf{\n{p}\n{q}}$ }
\RightLabel{$p$}
\BinaryInfC{ $\mathbf{\n{q}}$ }
\RightLabel{$q$}
\BinaryInfC{$\mathbf{r}$}
\end{prooftree}
\end{minipage}
&
(5)
\end{tabular}
\medskip \\

\nin{}$S2$ has now exposed a new redundancy involving
the variable $q$. The proof can be readily simplified by means of
an application of $R2'$:

\medskip
\begin{tabular}{cr}
\begin{minipage}{.8\textwidth}
\scriptsize
\begin{prooftree}
\AxiomC{ $qr$ }
\AxiomC{ $\mathbf{p\n{q}}$ }
\AxiomC{ $\mathbf{pq}$ }
\RightLabel{$q$}
\BinaryInfC{ $\mathbf{p}$ }
\AxiomC{ $\mathbf{\n{p}\n{q}}$ }
\RightLabel{$p$}
\BinaryInfC{ $\mathbf{\n{q}}$ }
\RightLabel{$q$}
\BinaryInfC{$r$}
\end{prooftree}
\end{minipage}
&
(6)
\end{tabular}
\medskip \\

\medskip
\begin{tabular}{cr}
\begin{minipage}{.8\textwidth}
\scriptsize
\begin{prooftree}
\AxiomC{ $qr$ }
\AxiomC{ $\mathbf{p\n{q}}$ }
\AxiomC{ $\mathbf{\n{p}\n{q}}$ }
\RightLabel{$p$}
\BinaryInfC{ $\mathbf{\n{q}}$ }
\RightLabel{$q$}
\BinaryInfC{$r$}
\end{prooftree}
\end{minipage}
&
(7)
\end{tabular}
\medskip \\

\nin{}As discussed in \S\ref{subsec:meta}, the rewriting framework defined by our rules 
allows the flexibility of choosing a strategy
and a termination criterion for their application. 

A simple strategy is to eagerly apply the $R$ rules until
possible, shuffle the proof by means of $S2$ with the purpose of 
disclosing other redundancies, and then apply the $R$ rules again, in an iterative fashion. 
However there is usually a very large number of contexts where $S2$ could be applied, and
it is computationally expensive to predict whether one or a chain of $S2$ applications 
would eventually lead to the creation of contexts for an $R$ rule.

For efficiency reasons, we rely on the meta-algorithm described in Alg.~\ref{alg:rectraloop}, for a particular
instantiation of the ApplyRule method. 
Alg.~\ref{alg:rectraloop} does a single traversal of the proof,
performing shuffling and compression; it is run multiple times, setting a number of traversals to perform and a timeout as termination criteria
 (whichever is reached first).
The resulting regularization procedure is \emph{ReduceAndExpose}, listed
as Alg.~\ref{alg:rectra}. 

\begin{algorithm}[!ht]
\SetAlgoNoLine
\LinesNumbered
\DontPrintSemicolon
\KwIn{A legal proof, \emph{timelimit}: timeout, \emph{numtrav}: number of transformation traversals, an instantiation of \emph{ApplyRule}}
\KwOut{A legal proof}
\Begin{
  \For{i=1 \KwTo numtrav}
  {
    TransformAndReconstruct(\emph{ApplyRule})\;
        \uIf{timelimit is reached}{break\;}
  }
}
\caption{ReduceAndExpose.}
\label{alg:rectra}
\end{algorithm}
\subsubsection{The RecyclePivots Approach}
\label{sec:recpiv}

The RecyclePivots algorithm was introduced in~\cite{BFHSS08} as a linear-time technique to perform a partial
regularization of resolution proofs. 

RecyclePivots is based on analyzing the paths of a proof, focusing on the pivots involved 
in the resolution steps; if a pivot is resolved upon more than once on a path (which implies that 
the pivot variable is introduced and then removed multiple times), the resolution step closest to the root is kept, 
while the others are simplified away.

We illustrate this approach by means of an example.
Consider the leftmost path of proof (1). 
Variable $p$ is used twice as pivot. The topmost resolution step is 
redundant as it resolves upon $p$, which is reintroduced in 
a subsequent step (curly brackets denote the set $RL$ of removable literals, see later).

\medskip
\begin{tabular}{cr}
\begin{minipage}{.8\textwidth}
\begin{prooftree}
\scriptsize
\AxiomC{ $\mathbf{pq}$ }
\AxiomC{ $\mathbf{\n{p}o}$ }
\RightLabel{$p$}
\BinaryInfC{ $\mathbf{qo} \ \{\n{p},\n{q}\}$ }
\AxiomC{ $\mathbf{p\n{q}}$ }
\RightLabel{$q$}
\BinaryInfC{ $\mathbf{po} \ \{\n{p}\}$ }
\AxiomC{ $qo$ }
\AxiomC{ $\n{p}\n{q}$ }
\RightLabel{$q$}
\BinaryInfC{ $\n{p}o$ }
\RightLabel{$p$}
\BinaryInfC{$o$}
\end{prooftree}
\end{minipage}
&
(1)
\end{tabular}
\medskip \\

Regularization can be achieved by eliminating the topmost
resolution step and by adjusting the proof accordingly. 
The resulting proof is shown below.

\medskip
\begin{tabular}{cr}
\begin{minipage}{.8\textwidth}
\begin{prooftree}
\scriptsize
\AxiomC{ $\mathbf{pq}$ }
\AxiomC{ $\mathbf{p\n{q}}$ }
\RightLabel{$q$}
\BinaryInfC{ $\mathbf{p}$ }
\AxiomC{ $qo$ }
\AxiomC{ $\n{p}\n{q}$ }
\RightLabel{$q$}
\BinaryInfC{ $\n{p}o$ }
\RightLabel{$p$}
\BinaryInfC{$o$}
\end{prooftree}
\end{minipage}
&
(2)
\end{tabular}
\medskip \\

Alg.~\ref{alg:recpiv} shows the recursive version of RecyclePivots (RP in the following).
It is based on a depth-first visit of the proof, from the root to the
leaves. 
It starts from the global root, having as input a set of \emph{removable literals RL} (initially empty). 
The removable literals are essentially the (partial) collection of pivot literals encountered 
during the bottom-up exploration of a path. If the pivot 
variable of a resolution step under consideration is in $RL$ (lines 15 and 18), 
then the resolution step is redundant and one of the antecedents may be removed 
from the proof. The resulting proof is illegal and has to be 
reconstructed into a legal one, which can be done in linear time, as shown in~\cite{BFHSS08}. 

Note that in the case of resolution proof trees, the outcome of the algorithm
is a regular proof. For arbitrary resolution proof DAGs  the algorithm is executed
in a limited form (when nodes with multiple children are detected) precisely by resetting $RL$
(line 10); therefore the result is not necessarily a regular proof.

\begin{algorithm}[!ht]
\SetAlgoNoLine
\LinesNumbered
\DontPrintSemicolon
\KwIn{A node $n$, a set of removable literals $RL$}
\Begin{
\uIf{$n$ is visited}{return\;}
\uElse{
	Mark $n$ as visited\;
	\uIf{$n$ is a leaf}{return\;}
	\uElse{
		\uIf{$n$ has more than one child}{$RL \leftarrow \emptyset$\;}
		$p \leftarrow piv(n)$\; 
		\uIf{$p \notin RL$ {and} $\n{p} \notin RL$} 
			{
			RecyclePivots($n^{+}$,$RL \cup \{\n{p}\}$)\;
			RecyclePivots($n^{-}$,$RL \cup \{p\}$)\;
			}
		\uElseIf{$p \in RL$} 
			{
			$n^{+} \leftarrow null$\;
			RecyclePivots($n^{-}$,$RL$)\;
			}
		\uElseIf{$\n{p} \in RL$} 
			{
			$n^{-} \leftarrow null$\;
			RecyclePivots($n^{+}$,$RL$)\;
			}
		}
	}
}
\caption{RecyclePivots(n,RL).}
\label{alg:recpiv}
\end{algorithm}

\subsubsection{RecyclePivotsWithIntersection}
\label{subsub:rpi}

The aforementioned limitation is due to the same circumstance that restricts the application of rules in the Local Transformation Framework, as discussed in \S\ref{subsec:ruledags}.
The set of removable literals of a node is computed for a particular path from the root to the node (which is enough in presence of proof trees),
but does not take into account the existence of other possible paths to that node.
Thus, suppose a node $n$ with  pivot $p$ is replaced by one of its parents (let us say $n^+$) during the reconstruction phase, and $C(n^+) \nsubseteq C(n)$; 
then, it might happen that
some of the literals in $C(n^+) \setminus C(n)$ are not resolved upon along \emph{all} paths from $n$ to the root, and are thus propagated to the root, making the proof illegal. 

In order to address this issue, the authors of~\cite{FMP11} extend RP by proposing RecyclePivotsWithIntersection (RPI),
an iterative version of which is illustrated in Alg.~\ref{alg:recpivint}.
RPI refines RP  by keeping track for each node $n$ of the set of pivot literals $RL(n)$ which get resolved upon 
along \emph{all} paths from $n$ to the root. 

\begin{figure}[!ht]
\centering
\begin{tabular}{| p{0.3\textwidth} |  p{0.3\textwidth} | p{0.28\textwidth} |}
\hline
\begin{minipage}{0.3\textwidth}
\centering
{\scriptsize
\begin{align*}
RL(n) = \bigcap (RL(m_i) \cup \{q_i\}) \\ 
q_i = \left\{\begin{array}{ll}
p_i \quad  \text{ if } \n{p_i} \in C(n) \\ 
 \n{p_i} \quad \text{ if } p_i \in C(n)
\end{array}\right.
\end{align*}
}
\end{minipage}
&
\begin{minipage}{0.3\textwidth}
\centering
{\scriptsize
\begin{align*}
RL(n) = (RL(m) \cup \{q\}) \\ 
q = \left\{\begin{array}{ll}
p \quad  \text{ if } \n{p} \in C(n) \\ 
 \n{p} \quad \text{ if } p \in C(n)
\end{array}\right.
\end{align*}
}
\end{minipage}
&
\begin{minipage}{0.28\textwidth}
\centering
{\scriptsize 
\begin{align*}
RL(n) = \bigcup \{q_i\} \\ 
q_i = \left\{\begin{array}{ll}
p_i \quad  \text{ if } \n{p_i} \in C(n) \\ 
 \n{p_i} \quad \text{ if } p_i \in C(n)
\end{array}\right.
\end{align*}
}
\end{minipage}
\\
\begin{minipage}{0.3\textwidth}
\centering
\begin{tikzpicture}[->,every node/.style={font=\sffamily\small}]
  \node (1) {$n$};
  \node (2) [left of=1] {};
  \node (3) [below of=1] {$\cdots$};
  \node (4) [right of=1] {};
  \node (5) [below of=2] {$m_1$};
  \node (6) [below of=4] {$m_k$};

  \path[every node/.style={font=\sffamily\scriptsize}]
    (1) edge node [left] {$p_1$} (5)
    (1) edge node [right] {$p_k$} (6);
\end{tikzpicture}
\end{minipage}
&
\begin{minipage}{0.3\textwidth}
\centering
\begin{tikzpicture}[->,every node/.style={font=\sffamily\small}]
  \node (1) {$n$};
  \node (3) [below of=1] {$m$};
  \path[every node/.style={font=\sffamily\scriptsize}]
    (1) edge node [left] {$p$} (3);
\end{tikzpicture}
\end{minipage}
&
\begin{minipage}{0.28\textwidth}
\centering
\begin{tikzpicture}[->,every node/.style={font=\sffamily\small}]
  \node (1) {$n$};
\end{tikzpicture}
\end{minipage}\\
& &\\
\hline
\end{tabular}
\caption{Computation of $RL$ in RecyclePivotsWithIntersection.}
\label{fig:rlrpi}
\end{figure}
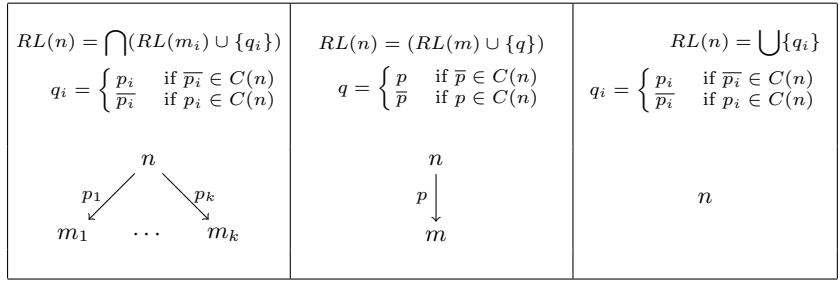

The computation of $RL$  in the two approaches is represented in Fig.~\ref{fig:rlrp} and Fig.~\ref{fig:rlrpi}.
RP and RPI behave in the same way whenever a node $n$ has only one child. In case $n$ has no children, i.e., it is the root, RPI takes into
account the possibility for the root to be an arbitrary clause (rather than only $\bot$, as in refutations) and sets $RL$ to include all  variables of $C(n)$;
it is equivalent to having a path from $n$ to $\bot$ where all variables of $C(n)$ are resolved upon.
The major difference between RP and RPI is in the way a node $n$ with multiple children is handled: RP sets $RL(n)$ to $\emptyset$, while
RPI sets $RL(n)$ to the intersection $\bigcap (RL(m_i) \cup q_i)$ of the removable literals sets of its children, augmented with the pivots of the resolution
steps of which the children are resolvents.

\begin{figure}[!ht]
\centering
\begin{tabular}{| p{0.3\textwidth} |  p{0.3\textwidth} | p{0.28\textwidth} |}
\hline
\begin{minipage}{0.3\textwidth}
\centering
{\scriptsize 
\begin{align*}
RL(n) = \emptyset
\end{align*}
}
\end{minipage}
&
\begin{minipage}{0.3\textwidth}
\centering
{\scriptsize
\begin{align*}
RL(n) = (RL(m) \cup \{q\}) \\ 
q = \left\{\begin{array}{ll}
p \quad  \text{ if } \n{p} \in C(n) \\ 
 \n{p} \quad \text{ if } p \in C(n)
\end{array}\right.
\end{align*}
}
\end{minipage}
&
\begin{minipage}{0.28\textwidth}
\centering
{\scriptsize 
\begin{align*}
RL(n) = \emptyset
\end{align*}
}
\end{minipage}
\\
\begin{minipage}{0.3\textwidth}
\centering
\begin{tikzpicture}[->,every node/.style={font=\sffamily\small}]
  \node (1) {$n$};
  \node (2) [left of=1] {};
  \node (3) [below of=1] {$\cdots$};
  \node (4) [right of=1] {};
  \node (5) [below of=2] {$m_1$};
  \node (6) [below of=4] {$m_k$};

  \path[every node/.style={font=\sffamily\scriptsize}]
    (1) edge node [left] {$p_1$} (5)
    (1) edge node [right] {$p_k$} (6);
\end{tikzpicture}
\end{minipage}
&
\begin{minipage}{0.3\textwidth}
\centering
\begin{tikzpicture}[->,every node/.style={font=\sffamily\small}]
  \node (1) {$n$};
  \node (3) [below of=1] {$m$};
  \path[every node/.style={font=\sffamily\scriptsize}]
    (1) edge node [left] {$p$} (3);
\end{tikzpicture}
\end{minipage}
&
\begin{minipage}{0.28\textwidth}
\centering
\begin{tikzpicture}[->,every node/.style={font=\sffamily\small}]
  \node (1) {$n$};
\end{tikzpicture}
\end{minipage}\\
& & \\
\hline
\end{tabular}
\caption{Computation of $RL$ in RecyclePivots.}
\label{fig:rlrp}
\end{figure}
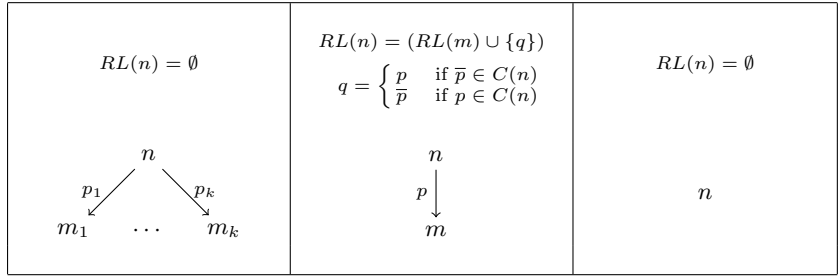

\begin{algorithm}[!ht]
\SetAlgoNoLine
\LinesNumbered
\DontPrintSemicolon
\KwIn{A legal proof}
\KwIn{A proof to be reconstructed}
\KwData{$TS$: nodes topological sorting vector, $RL$: vector of sets of removable literals }
\Begin{
  $TS \leftarrow$ topological\_{}sorting\_bottom\_up(proof)\;
  \ForEach{$n \in TS$}{
    \uIf{$n$ {is not a leaf}}{
		\uIf{$n$ {is the root}}
		{ $RL(n)\leftarrow \{\n{p_i}\}_{p_i \in C(n)}$\; }
		\uElse{
		$p \leftarrow piv(n)$\; 
		\uIf{$p \in RL(n)$} 
			{
			$n^{+} \leftarrow null$\;
			\uIf{$n^-$ {not seen yet}}
			{
			$RL(n^{-}) \leftarrow RL(n)$\;
			Mark $n^-$ as seen\;
			}
			\lElse
			{ $RL(n^{-}) \leftarrow RL(n^-) \cap RL(n)$\; }
			}
		\uElseIf{$\n{p} \in RL(n)$} 
			{
			$n^{-} \leftarrow null$\;
			\uIf{$n^+$ {not seen yet}}
			{
			$RL(n^{+}) \leftarrow RL(n)$\;
			Mark $n^+$ as seen\;
			}
			\lElse
			{ $RL(n^{+}) \leftarrow RL(n^+) \cap RL(n)$\; }
			}
		\uElseIf{$p \notin RL(n)$ {and} $\n{p} \notin RL(n)$} 
			{
			
			\uIf{$n^-$ {not seen yet}}
			{
			$RL(n^{-}) \leftarrow (RL(n) \cup \{p\})$\;
			Mark $n^-$ as seen\;
			}
			\lElse
			{ $RL(n^{-}) \leftarrow RL(n^-) \cap (RL(n) \cup \{p\})$\; }
			
			\uIf{$n^+$ {not seen yet}}
			{
			$RL(n^{+}) \leftarrow (RL(n) \cup \{\n{p}\})$\;
			Mark $n^+$ as seen\;
			}
			\lElse
			{ $RL(n^{+}) \leftarrow RL(n^+) \cap (RL(n) \cup \{\n{p}\})$\; }
			
			}
		}
		}
	}
}
\caption{RecyclePivotsWithIntersection.}
\label{alg:recpivint}
\end{algorithm}

RPI starts in Alg.~\ref{alg:recpivint} by computing a topological sorting of the nodes (line 2), from the root to the leaves. 
$RL(root)$ is computed as the set  of literals in the root clause; for any other node $n$, $RL(n)$
is initialized and then iteratively refined each time one of its children is visited. 
Similarly to RecyclePivots, whenever visiting an inner node $n$, if $piv(n)$ appears in $RL(n)$ then the resolution step is redundant
and can be simplified away (lines 9-14, 15-20); in that case, $RL(n)$ is propagated to a parent of $n$  without the addition of $piv(n)$.

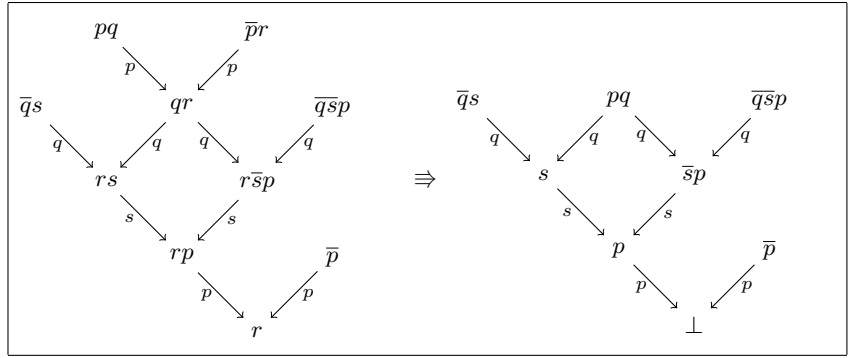
\begin{figure}[!ht]
\centering
\renewcommand{\arraystretch}{0.34}
\begin{tabular}{|p{0.43\textwidth} p{0.03\textwidth} p{0.43\textwidth}|}
\hline
& & \\
\begin{minipage}{0.43\textwidth}
\begin{tikzpicture}[->,every node/.style={font=\sffamily\small}]
  \node (1) {$pq$};
  \node (2) [right of=1] {};
  \node (3) [right of=2] {$\n{p}r$};
  \node (4) [below of=2] {$qr$};
  \node (5) [right of=4] {};
  \node (6) [right of=5] {$\n{q}\n{s}p$};
  \node (7) [below of=5] { $r\n{s}p$};
  \node (8) [left of=4] {};
  \node (9) [left of=8] {$\n{q}s$};
  \node (10) [below of=8] {$rs$};
  \node (11) [left of=7] {};
  \node (12) [below of=11] {$rp$};
  \node (13) [right of=12] {};
  \node (14) [right of=13] {$\n{p}$};
  \node (15) [below of=13] {$r$};

  \path[every node/.style={font=\sffamily\scriptsize}]
    (1) edge node [left] {$p$} (4)
    (3) edge node [right] {$p$} (4)
    (9) edge node [left] {$q$} (10)
    (4) edge node [right] {$q$} (10)
    (4) edge node [left] {$q$} (7)
    (6) edge node [right] {$q$} (7)
    (7) edge node [right] {$s$} (12)
    (10) edge node [left] {$s$} (12)
    (14) edge node [right] {$p$} (15)
    (12) edge node [left] {$p$} (15);
\end{tikzpicture}
\end{minipage}
&
$\Rrightarrow$ 
&
\begin{minipage}{0.43\textwidth}
\begin{tikzpicture}[->,every node/.style={font=\sffamily\small}]
  \node (1) {};
  \node (2) [right of=1] {};
  \node (3) [right of=2] {};
  \node (4) [below of=2] {$pq$};
  \node (5) [right of=4] {};
  \node (6) [right of=5] {$\n{q}\n{s}p$};
  \node (7) [below of=5] { $\n{s}p$};
  \node (8) [left of=4] {};
  \node (9) [left of=8] {$\n{q}s$};
  \node (10) [below of=8] {$s$};
  \node (11) [left of=7] {};
  \node (12) [below of=11] {$p$};
  \node (13) [right of=12] {};
  \node (14) [right of=13] {$\n{p}$};
  \node (15) [below of=13] {$\bot$};

  \path[every node/.style={font=\sffamily\scriptsize}]
    (9) edge node [left] {$q$} (10)
    (4) edge node [right] {$q$} (10)
    (4) edge node [left] {$q$} (7)
    (6) edge node [right] {$q$} (7)
    (7) edge node [right] {$s$} (12)
    (10) edge node [left] {$s$} (12)
    (14) edge node [right] {$p$} (15)
    (12) edge node [left] {$p$} (15);
\end{tikzpicture}
\end{minipage}\\
& & \\
\hline
\end{tabular}
\caption{Compression of a proof by means of RecyclePivotsWithIntersection.}
\label{fig:rpiex}
\end{figure}

Fig.~\ref{fig:rpiex} shows the effect of RPI on a small proof where RP cannot achieve any compression:
RP sets $RL(qr)=\emptyset$ since $qr$ has two children, while RPI sets $RL(qr)=\{\n{r},\n{p},\n{q}\}$ 
and consequently simplifies the uppermost resolution step, since it is able to detect that $p$ is
resolved upon along both paths from $qr$ to the root.

\subsubsection{RecyclePivots and the Local Transformation Framework}
\label{sec:combined}

RecyclePivots (as well as its refinement RecyclePivotsWithIntersection) 
and ReduceAndExpose both aim at compressing
a proof by identifying and removing pivot redundancies along paths from the root
to the leaves. The main difference between the two approaches is that RecyclePivots
operates on a {\em global perspective} without changing the topology of the proof (i.e., no shuffling), 
while ReduceAndExpose operates on {\em local contexts} and allows
the topology to change. Both approaches have advantages and disadvantages.

Operating on a global perspective without modifying the topology allows a one-pass
visit and compression of the proof. Maintaining a fixed topology, however,
may prevent the disclosure of hidden redundancies. For instance the application of
RecyclePivots to the example of~\S\ref{sec:traalgo} would have stopped to step (3), since
no more redundant pivots can be found along a path (the proof is regular). 
The local contexts instead have to be gathered and considered multiple times. On the 
other hand, the ability of ReduceAndExpose to change the topology allows
more redundancies to be exposed.

Another advantage of RecyclePivots is that it can eliminate redundancies that are separated
by many resolution steps. The $R$ rewriting rules instead are applicable only when there is
a reintroduction of a certain variable immediately after a resolution step upon it. 
Such configurations, when not present in the proof, can be produced by means of 
applications of the $S2$ rule. 

The ability of the Local Transformation Framework to disclose redundancies and the effectiveness
of RecyclePivots at removing them can be combined in a simple hybrid approach, shown in Alg.~\ref{alg:combalg1}.

\begin{algorithm}[!ht]
\SetAlgoNoLine
\LinesNumbered
\DontPrintSemicolon
\KwIn{A legal proof,
      $numloop$: number of global iterations,
      $numtrav$: number of transformation traversals for each global iteration,
      $timelimit$: timeout,
      an instantiation of \emph{ApplyRule}}
\KwOut{A  legal proof}
\Begin{
  $timeslot=timelimit/numloop$\;
  \For{i=1 \KwTo $numloop$}
  {
    RecyclePivots($root$,$\emptyset$)\;
    // $RPtime$ is the time taken by RecyclePivots in the last call\;
    ReduceAndExpose($timeslot-RPtime$,$numtrav$,$ApplyRule$)\;
  }
}
\caption{RP + RE.}
\label{alg:combalg1}
\end{algorithm}

The algorithm takes as input an overall time limit, a number of \emph{global iterations} and a number of transformation traversals for ReduceAndExpose.
The time limit and the amount of global iterations determine the execution time available to ReduceAndExpose during each iteration.
ReduceAndExpose and RecyclePivots are run one after the other by Alg.~\ref{alg:combalg1}, alternately modifying the topology to expose redundancies
and simplifying them away.

A similar, but more efficient algorithm can be obtained by simply replacing the call to RecyclePivots with a call to RecyclePivotsWithIntersection.

\subsection{Proof Compactness}

The focus of this section is the notion of compactness as introduced in \S\ref{subsec:red}: a proof is compact whenever different nodes
are labeled with different clauses, that is $C(n_i) = C(n_j) \implies i = j$ for any $i,j$.
We first present an algorithm to address redundancies related to the presence of multiple occurrences of a same unit clause in a proof.
Then we illustrate a technique based on a form of structural hashing, which makes a proof more compact
by identifying and merging nodes having exactly the same pair of parents.
We conclude by showing how to combine these procedures with the Local Transformation Framework.

\subsubsection{Unit Clauses-Based Simplification}
\label{subsub:unit}

The simplification of a proof by exploiting the presence of unit clauses has already been addressed in the literature in~\cite{FMP11}
and~\cite{BFHSS08}. The two works pursue different goals. The \emph{RecycleUnits} algorithm from~\cite{BFHSS08} uses learned unit clauses to rewrite subproofs  
that were derived before learning them. On the other hand, the \emph{LowerUnits} algorithm from~\cite{FMP11} collects unit clauses and reinserts them  
at the level of the global root, thus removing redundancies due to multiple resolution steps on the same unit clauses. 

Following the idea of~\cite{FMP11}, we present \emph{PushdownUnits}, listed as Alg.~\ref{alg:simp}.
First, the algorithm traverses a proof in a top-down manner, detaching and collecting subproofs rooted in unit clauses,
while at the same time reconstructing the proof to keep it legal (based on the schema of Alg.~\ref{alg:rectraloop});
then, (some of) these subproofs are attached back at the end of the proof, adding new resolution steps.
PushdownUnits improves over LowerUnits by performing unit collection and proof reconstruction in a single pass.

\begin{algorithm}[!ht] 
\SetAlgoNoLine
\LinesNumbered
\DontPrintSemicolon
\KwIn{A legal proof}
\KwOut{A legal proof}
\KwData{$TS$: nodes topological sorting vector, $CU$: collected units set, $EL$: set of extra literals appearing in the global root}
\Begin{
  $TS \leftarrow$ topological\_{}sorting\_top\_down(proof)\;
  $r \leftarrow $ global root\;
  \ForEach{$n \in TS$}{
    \uIf{$n$ {is not a leaf}}{
    $p \leftarrow piv(n)$\;
      \uIf{$\n{p} \in C(n^-)$ {and} $p \in C(n^+)$}{
	$C(n) \leftarrow Res_{p}(C(n^-),C(n^+))$\;
	\uIf{$C(n^+) = p$}{
	Substitute $n$ with $n^-$\;
	$CU \leftarrow CU \cup \{n^+\}$
	}
	\uElseIf{$C(n^-) = \n{p}$}{
	Substitute $n$ with $n^+$\;
	$CU \leftarrow CU \cup \{n^-\}$
	}
      }
      \uElseIf{$piv(n) \notin C(n^-)$ {and} $piv(n) \in C(n^+)$}{
	Substitute $n$ with $n^-$\;
      }
      \uElseIf{$piv(n) \in C(n^-)$ {and} $piv(n) \notin C(n^+)$}{
	Substitute $n$ with $n^+$\;
      }
      \uElseIf{$piv(n) \notin C(n^-)$ {and} $piv(n) \notin C(n^+)$}{
	Heuristically choose a parent, substitute $n$ with it\;
      }
    }
  }
    $EL \leftarrow$ extra literals of $C(r)$\;
    \ForEach{$m \in CU$}{
    $s \leftarrow C(m)$\;
    \uIf{$\n{s} \in EL$}{
    Add a new node $o$ s.t.  $C(o) = Res_{v(s)}(C(r),C(m))$\;
    $r \leftarrow o$\;
    }
    }
}
\caption{PushdownUnits.}
\label{alg:simp}
\end{algorithm}

The algorithm works as follows. The proof is traversed according to a topological order.
When a node $n$ is visited s.t. $C(n)$ is the resolvent of a sound resolution step with pivot $p$, its parents are examined.
Assume $n^+$ is a unit clause, that is $C(n^+)=p$; then $n$ is replaced by the other parent $n^-$ and $n^+$ is added to the set of unit clauses $CU$.

This transformation phase might add extra literals $EL$ to the original global root $r$; if this is the case, the necessary resolution steps to make the proof legal are added at the end, 
starting from $r$.
The nodes previously collected are taken into account one by one; for each $m$, if $C(m)=s$ and $\n{s}$ is one of the extra literals $EL$, then a new resolution
step is added and its resolvent becomes the new root. 

Note that not necessarily all these nodes will be added back to the proof. 
Multiple nodes might be labeled by the same literal, in which case the correspondent variable will be used only once as pivot. Also, a collected literal which was an antecedent
of some resolution step might have been anyway resolved upon again along all paths from that resolution step to the global root; 
if so, it does not appear in the set of extra literals. The subproofs rooted in these unnecessary nodes can be (partially) pruned away to further compress the proof.

\begin{figure}[!ht]
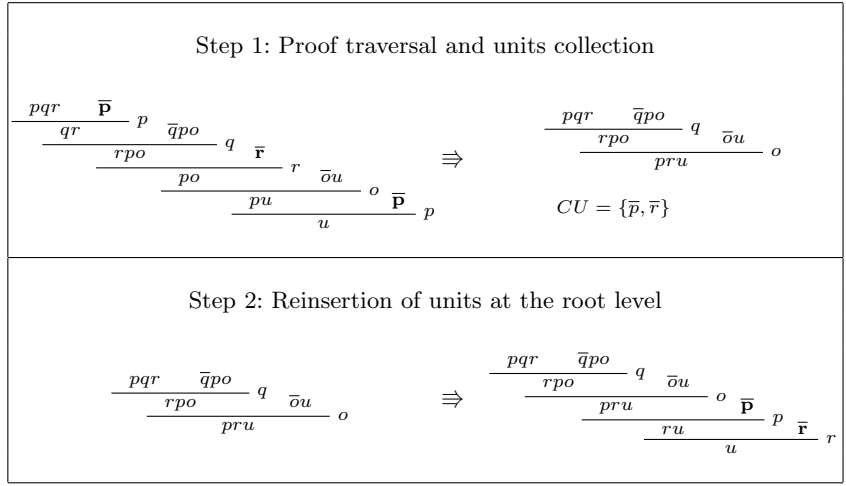

\centering
\begin{tabular}{|lcr|}
\hline
& & \\
\multicolumn{3}{|c|}{Step 1: Proof traversal and units collection} \\
& & \\
\begin{minipage}{.46\textwidth}
\scriptsize
\def\defaultHypSeparation{\hskip .01in}
\begin{prooftree}
\AxiomC{ $pqr$ }
\AxiomC{ $\mathbf{\n{p}}$ }
\RightLabel{$p$}
\BinaryInfC{ $qr$ }
\AxiomC{ $\n{q}po$ }
\RightLabel{$q$}
\BinaryInfC{ $rpo$ }
\AxiomC{ $\mathbf{\n{r}}$ }
\RightLabel{$r$}
\BinaryInfC{ $po$ }
\AxiomC{ $\n{o}u$ }
\RightLabel{$o$}
\BinaryInfC{ $pu$ }
\AxiomC{$\mathbf{\n{p}}$}
\RightLabel{$p$}
\BinaryInfC{ $u$ }
\end{prooftree}
\end{minipage}
& $\Rrightarrow$ &
\begin{minipage}{.40\textwidth}
\scriptsize
\def\defaultHypSeparation{\hskip .02in}
\begin{prooftree}
\AxiomC{ $pqr$ }
\AxiomC{ $\n{q}po$ }
\RightLabel{$q$}
\BinaryInfC{ $rpo$ }
\AxiomC{ $\n{o}u$ }
\RightLabel{$o$}
\BinaryInfC{ $pru$ }
\end{prooftree}
$\qquad \qquad CU=\{\n{p},\n{r}\}$
\end{minipage} \\
& & \\
\hline
& & \\
\multicolumn{3}{|c|}{Step 2: Reinsertion of units at the root level} \\
& & \\
\begin{minipage}{.46\textwidth}
\scriptsize
\def\defaultHypSeparation{\hskip .02in}
\begin{prooftree}
\AxiomC{ $pqr$ }
\AxiomC{ $\n{q}po$ }
\RightLabel{$q$}
\BinaryInfC{ $rpo$ }
\AxiomC{ $\n{o}u$ }
\RightLabel{$o$}
\BinaryInfC{ $pru$ }
\end{prooftree}
\end{minipage}
& $\Rrightarrow$ &
\begin{minipage}{.40\textwidth}
\scriptsize
\def\defaultHypSeparation{\hskip .02in}
\begin{prooftree}
\AxiomC{ $pqr$ }
\AxiomC{ $\n{q}po$ }
\RightLabel{$q$}
\BinaryInfC{ $rpo$ }
\AxiomC{ $\n{o}u$ }
\RightLabel{$o$}
\BinaryInfC{ $pru$ }
\AxiomC{$\mathbf{\n{p}}$}
\RightLabel{$p$}
\BinaryInfC{ $ru$ }
\AxiomC{$\mathbf{\n{r}}$}
\RightLabel{$r$}
\BinaryInfC{ $u$ }
\end{prooftree}
\end{minipage} \\
& & \\
\hline
\end{tabular}
\caption{Example of application of PushdownUnits. Note that the lowest occurrence of $\n{p}$ is not added back to the proof.}
\label{fig:simpun}
\end{figure}

\subsubsection{Structural Hashing}
\label{subsub:sh}

The work of~\cite{C10} proposes an algorithm based on a form of \emph{structural hashing}; it explicitly takes into account how resolution proofs
are obtained in CDCL SAT-solvers from a sequence of subproofs deriving learnt clauses, and keeps a hash map
which stores for each derived clause its pair of antecedents.
While building the global proof from the sequence of subproofs, whenever a clause would be added,
if its pair of antecedents is already in the hash map, then the existing clause is used.

Taking inspiration from the idea at the base of this technique, we present a post-processing compression algorithm, \emph{StructuralHashing}, which aims at improving
the compactness of a proof.
StructuralHashing is illustrated in Alg.~\ref{alg:structhash}.

\begin{algorithm}[!ht] 
\SetAlgoNoLine
\LinesNumbered
\DontPrintSemicolon
\KwIn{A legal proof}
\KwOut{A legal proof}
\KwData{$TS$: nodes topological sorting vector, $HM$: hash map associating a node to its pair of parents}
\Begin{
  $TS \leftarrow$ topological\_{}sorting\_top\_down(proof)\;
  \ForEach{$n \in TS$}{
    \uIf{$n$ {is not a leaf}}{
      \uIf{$<n^{+},n^{-}> \in HM$}{
	$m  \leftarrow HM(<n^{+},n^{-}>)$\;
	Replace $n$ with $m$\;
	Assign $n$ children to $m$\;
      }
      \uElse{
      $HM(<n^{+},n^{-}>) \leftarrow n$
      }
    }
  }
}
\caption{StructuralHashing.}
\label{alg:structhash}
\end{algorithm}

The proof is traversed in topological order. 
When a node $n$ is visited, the algorithm first checks whether its antecedents are already in the hash map; if so,  another node $m$ with
the same parents has been seen before. In that case, $n$ is replaced by $m$ and the children of $n$ are assigned to $m$.
The use of a topological visit guarantees the soundness of the algorithm: it is safe to replace the subproof rooted in $n$ with that rooted in $m$
since either ($i$) $m$ is an ancestor of $n$ (and the subproof rooted in $m$ is contained in the subproof rooted in $n$) or ($ii$) $m$ and $n$ are not on a same path to the global root,
so $m$ is not involved in the derivation of $n$.

Note that StructuralHashing does not guarantee a completely compact proof;
if two nodes $n_1,n_2$ have the same parents, then $C(n_1)=C(n_2)$, but the converse is not necessarily true.
A complete but more computationally expensive technique might consist in employing a hash map to associate clauses with nodes (rather than
pairs of nodes with nodes as done in StructuralHashing), 
based on a function that derives map keys from the clauses content; an implementation of this technique can be found in~\cite{skeptic}.

\subsubsection{StructuralHashing and the Local Transformation Framework}

StructuralHashing is a one-pass compression technique, like RecyclePivots and RecyclePivotsWithIntersection.
Nevertheless, it is still possible to exploit the Local Transformation Framework in order to disclose new redundancies and remove them, 
in an iterative manner. We illustrate this approach in Alg.~\ref{alg:combalg2}.
\begin{algorithm}[!ht]
\SetAlgoNoLine
\LinesNumbered
\DontPrintSemicolon
\KwIn{A legal proof,
      $numloop$: number of global iterations,
      $numtrav$: number of transformation traversals for each global iteration,
      $timelimit$: timeout,
      an instantiation of \emph{ApplyRule}}
\KwOut{A legal proof}
\Begin{
  $timeslot=timelimit/numloop$\;
  \For{i=1 \KwTo $numloop$}
  {
    StructuralHashing()\;
    // $SHtime$ is the time taken by StructuralHashing in the last call\;
    ReduceAndExpose($timeslot-SHtime$,$numtrav$,$ApplyRule$)\;
  }
}
\caption{SH + RE.}
\label{alg:combalg2}
\end{algorithm}

\subsubsection{A Synergic Algorithm}
\label{subsubsub:syn}

It is possible to combine the compression techniques illustrated so far as shown in Alg.~\ref{alg:combalg3}, exploiting their
individual features for a synergistic effect. The combined approach executes the algorithms sequentially for a given number
of \emph{global iterations}.
Note that PushdownUnits is kept outside of the loop: in our experience, SH, RPI and RE are unlikely to introduce 
unit clauses in the proofs, thus for efficiency PushdownUnits is run only once before the main loop. 

\begin{algorithm}[!ht]
\SetAlgoNoLine
\LinesNumbered
\DontPrintSemicolon
\KwIn{A legal proof,
      $numloop$: number of global iterations,
      $numtrav$: number of transformation traversals for each global iteration,
      $timelimit$: timeout,
      an instantiation of \emph{ApplyRule}}
\KwOut{A legal proof}
\Begin{
  $timeslot=timelimit/numloop$\;
  PushdownUnits()\;
  \For{i=1 \KwTo $numloop$}
  {
    StructuralHashing()\;
    RecyclePivotsWithIntersection()\;
    // $SHtime$ and $RPItime$ are the time taken by StructuralHashing \newline
    // and  RecyclePivotsWithIntersection in the last call\;
    ReduceAndExpose($timeslot-SHtime-RPItime$,$numtrav$,$ApplyRule$)\;
  }
}
\caption{PU + SH + RPI + RE.}
\label{alg:combalg3}
\end{algorithm}

The overall complexity of the combined algorithm is parametric in the number of global iterations and actual transformation traversals (also depending on the specified time limit).

PushdownUnits performs a topological visit of the proof, collecting unit clauses and adding them back at the level of the global root;
the complexity is $O(|V|+|E|)$, linear in the size of the resolution proof DAG. 

Complexity is  $O(|V|+|E|)$ also for StructuralHashing,
which traverses the proof once, making use of an hash table to detect the existence of multiple nodes with the same resolvents.

An iterative implementation of RecyclePivotsWithIntersection consists of a bottom-up scan of the proof, while computing the sets of removable literals and pruning branches, 
followed by a reconstruction phase; the complexity is again $O(|V|+|E|)$.

Each execution of TransformAndReconstruct, on which ReduceAndExpose is based, computes a topological sorting of the nodes and traverses the proof top-down applying rewriting rules. 
If $m$ transformation traversals are executed, the complexity of ReduceAndExpose is  $O(m(|V|+|E|))$.

Note that PushdownUnits, RecyclePivotsWithIntersection, TransformAndReconstruct also perform  operations at the level of clauses, checking the presence
of pivots, identifying rule contexts, updating resolvents. These operations depend on the width of the involved clauses; in practice, this value is very small compared to the 
proof size, and the complexity can be considered $O(|V|+|E|)$. 

Finally, if $n$ global iterations are carried out, the total complexity is $O(nm(|V|+|E|))$. 
There is a clear trade-off between efficiency and compression. The higher the value of $m$ is, the more redundancies are exposed and then removed;
in practice, however, especially in case of large proofs, a complexity higher than linear cannot be afforded, so the $nm$ factor should be kept
constant in the size of the proofs.

Some heuristics on the application of the local rules in conjunction with RecyclePivots, RecyclePivotsWithIntersection and StructuralHashing 
have been proved particularly successful: we refer the reader to \S\ref{subsec:expopensmt}, \S\ref{sec:compexp2} and \S\ref{sec:heuri} for details.

\subsection{Experiments on SMT Benchmarks}
\label{subsec:expopensmt}

As a first stage of experimentation, we carried out an evaluation of the three algorithms RecyclePivots (RP), ReduceAndExpose (RE), 
and their combination RP+RE. The algorithms were implemented inside the tool OpenSMT~\cite{BPS+10}, 
with proof-logging capabilities enabled. 

We experimented on the set of unsatisfiable benchmarks taken from the 
SMT-LIB~\cite{SMTLIB} from the categories QF\_{}UF, QF\_{}IDL, QF\_{}LRA, 
QF\_{}RDL. For these sets of benchmarks we noticed that the aforementioned  
compression techniques are very effective. We believe that the reason is connected with the fact 
that the introduction of theory lemmata in SMT is performed lazily: the delayed introduction
of clauses involved in the final proof may negatively impact the online proof construction
in the SAT-solver.

All the experiments were carried out on a 32-bit Ubuntu server featuring a Dual-Core
2GHz Opteron CPU and 4GB of memory; a timeout of 600 seconds and a memory 
threshold of 2GB (whatever is reached first) were put as limit to the executions.

\begin{table}[!ht] 
\caption{Results for SMT benchmarks. \#Bench reports the number 
         of benchmarks solved and processed within the time/memory constraints, RedNodes\% and
	 RedEdges\% report the average compression in the number of nodes and edges of the proof graphs,
         and RedCore\% reports the average compression in the unsatisfiable core size. TranTime
         is the average transformation time in seconds.}
\medskip
\centering
\begin{tabular}{|p{0.7cm} || p{0.7cm} | p{0.7cm} ||  p{0.7cm} | p{0.7cm} || p{0.7cm} | p{0.7cm} || p{0.7cm} | p{0.7cm} | p{0.7cm} | p{0.7cm} |}
\hline
 & \multicolumn{2}{|c|}{\scriptsize{\#Bench}} 
 & \multicolumn{2}{|c|}{\scriptsize{RedNodes\%}} 
 & \multicolumn{2}{|c|}{\scriptsize{RedEdges\%}} 
 & \multicolumn{2}{|c|}{\scriptsize{RedCore\%}} 
 & \multicolumn{2}{|c|}{\scriptsize{TranTime(s)}} \\
\hline
\hline
\scriptsize{RP} & \multicolumn{2}{|c|}{1370} 
   & \multicolumn{2}{|c|}{6.7} 
   & \multicolumn{2}{|c|}{7.5} 
   & \multicolumn{2}{|c|}{1.3} 
   & \multicolumn{2}{|c|}{1.7} \\
\hline
\end{tabular} 
\smallskip \\
(a)
\bigskip \\
\begin{tabular}{|p{0.7cm} || p{0.6cm} | p{0.7cm} || p{0.6cm} | p{0.7cm} || p{0.6cm} | p{0.7cm} || p{0.6cm} | p{0.7cm} || p{0.6cm} | p{0.7cm} |}
\hline
 & \multicolumn{2}{|c||}{\scriptsize{\#Bench}} 
 & \multicolumn{2}{|c||}{\scriptsize{RedNodes\%}} 
 & \multicolumn{2}{|c||}{\scriptsize{RedEdges\%}} 
 & \multicolumn{2}{|c||}{\scriptsize{RedCore\%}} 
 & \multicolumn{2}{|c|}{\scriptsize{TranTime(s)}} \\
\hline
\hline
\scriptsize{Ratio} & \scriptsize{RE} & $\!\!\!$\scriptsize{RP+RE} & \scriptsize{RE} & $\!\!\!$\scriptsize{RP+RE} & \scriptsize{RE} & $\!\!\!$\scriptsize{RP+RE} & \scriptsize{RE} & $\!\!\!$\scriptsize{RP+RE} & \scriptsize{RE} & $\!\!\!$\scriptsize{RP+RE}\\
\hline
\hline
0.01 & 1364 & 1366 & 2.7 & 8.9 & 3.8 & 10.7 & 0.2 & 1.4 & 3.5 & 3.4 \\
\hline
0.025 & 1363 & 1366 & 3.8 & 9.8 & 5.1 & 11.9  & 0.3 & 1.5 & 3.6 & 3.6 \\
\hline
0.05 & 1364 & 1366 & 4.9 & 10.7 & 6.5 & 13.0  & 0.4 & 1.6 & 4.3 & 4.1 \\
\hline
0.075 & 1363 & 1366 & 5.7 & 11.4 & 7.6 & 13.8 & 0.5 & 1.7 & 4.8 & 4.5 \\
\hline
0.1 & 1361 & 1364 & 6.2 & 11.8 & 8.3 & 14.4 & 0.6 & 1.7 & 5.3 & 5.0 \\
\hline
0.25 & 1357 & 1359  & 8.4 & 13.6 & 11.0 & 16.6  & 0.9 & 1.9 & 8.2 & 7.6 \\
\hline
0.5 & 1346 & 1348 & 10.4 & 15.0 & 13.3 & 18.4 & 1.1 & 2.0 & 12.1 & 11.5 \\
\hline
0.75 & 1339 & 1341 & 11.5 & 16.0 & 14.7  & 19.5 & 1.2 & 2.1 & 15.8 & 15.1\\
\hline
1 & 1335 & 1337 & 12.4 & 16.7 & 15.7 & 20.4 & 1.3 & 2.2 & 19.4 & 18.8\\
\hline
\end{tabular}
\smallskip \\
(b)
\label{tbl:resSMTallAVG}
\end{table}
\begin{table}[!ht]
\caption{Results for SMT benchmarks. MaxRedNodes\% and MaxRedEdges\% are the maximum compression 
	 of nodes and edges achieved by the algorithms in the suite on individual benchmarks.
         }
\medskip
\centering
\begin{tabular}{|p{0.8cm} || p{1.9cm} | p{1.9cm} || p{1.9cm} | p{1.9cm} || p{1.9cm} | p{1.9cm} |}
\hline
& \multicolumn{2}{|c|}{\scriptsize{MaxRedNodes\%}} & \multicolumn{2}{|c|}{\scriptsize{MaxRedEdges\%}} & \multicolumn{2}{|c|}{\scriptsize{MaxRedCore\%}} \\
\hline
\hline
\scriptsize{RP} & \multicolumn{2}{|c|}{65.1} & \multicolumn{2}{|c|}{68.9} & \multicolumn{2}{|c|}{39.1} \\
\hline
\end{tabular}
\smallskip \\
(a) 
\bigskip \\
\begin{tabular}{|c||c|c||c|c||c|c|}
\hline
& \multicolumn{2}{|c|}{\scriptsize{MaxRedNodes\%}} & \multicolumn{2}{|c|}{\scriptsize{MaxRedEdges\%}} & \multicolumn{2}{|c|}{\scriptsize{MaxRedCore\%}} \\
\hline
\hline
\scriptsize{Ratio} & \scriptsize{RE} & \scriptsize{RP+RE} & \scriptsize{RE} & \scriptsize{RP+RE} & \scriptsize{RE} & \scriptsize{RP+RE} \\
\hline
\hline
0.01  & 54.4 & 66.3 & 67.7 & 70.2 & 45.7 & 45.7\\
\hline
0.025 & 56.0 & 77.2 & 69.5 & 79.9 & 45.7 & 45.7\\
\hline
0.05  & 76.2 & 78.5 & 78.9 & 81.2 & 45.7 & 45.7\\
\hline
0.075 & 76.2 & 78.5 & 79.7 & 81.2 & 45.7 & 45.7\\
\hline
0.1   & 78.2 & 78.8 & 82.9 & 83.6 & 45.7 & 45.7\\
\hline
0.25  & 79.3 & 79.6 & 84.1 & 84.4 & 45.7 & 45.7\\
\hline
0.5   & 76.2 & 79.1 & 83.3 & 85.2 & 45.7 & 45.7\\
\hline
0.75  & 78.2 & 79.9 & 84.4 & 86.1 & 45.7 & 45.7\\
\hline
1     & 78.3 & 79.9 & 84.6 & 86.1 & 45.7 & 45.7\\
\hline
\end{tabular}
\smallskip \\
(b)
\label{tbl:resSMTallTOP}
\end{table}

The executions of RE and RP+RE are parameterized with a time threshold, which we set as a fraction of the time taken by
the solver to solve the benchmarks: more difficult instances are likely to produce larger proofs, and therefore more
time is necessary to achieve compression.
Notice that, regardless of the ratio, RE and RP+RE both perform at least one complete transformation loop, 
which could result in an execution time slightly higher than expected for low ratios and small proofs.

Table~\ref{tbl:resSMTallAVG} shows the average proof compression after the application of the 
algorithms\footnote{Full experimental data, as well as executables used 
in tests are available at\\ \url{http://verify.inf.usi.ch/sites/default/files/RPTCI2014.tar.gz}}.
Table~\ref{tbl:resSMTallAVG}a shows the compression obtained after the execution of RP. Table~\ref{tbl:resSMTallAVG}b
instead shows the compression obtained with RE and RP+RE parameterized with
a timeout (ratio $\cdot$ solving time). In the columns we report the compression in the number of nodes and edges, the compression
of the unsatisfiable core, and the actual transformation time.
Table~\ref{tbl:resSMTallTOP} is organized as Table~\ref{tbl:resSMTallAVG} except that it reports the best compression
values obtained over all the benchmarks suites.

On a single run RP clearly achieves the best results for compression 
with respect to transformation time. To get the same effect on average on nodes and edges, 
for example, RE needs about 5 seconds and a ratio transformation time/solving time equal to 0.1, 
while RP needs less than 2 seconds. As for core compression, the ratio must grow up to 1.
On the other hand, as already remarked, RP cannot be run more than once.

The combined approach RP+RE shows a performance which is indeed better than the other two algorithms 
taken individually. It is interesting to see that the global perspective adopted by RP gives an 
initial substantial advantage, which is slowly but constantly reduced as more and more time is 
dedicated to local transformations and simplifications.

Table \ref{tbl:resSMTallTOP}b displays some remarkable peaks of compression obtained with the RE and RP+RE approaches
on the best individual instances. Interestingly we noticed that in some benchmarks, like 
\emph{24.800.graph} of the QF\_IDL suite, RP does not achieve any compression, due to the
high amount of nodes with multiple resolvents present in its proof that forces RecyclePivots 
to keep resetting the removable literals set RL. RP+RE instead, even for a very small ratio (0.01), 
performs remarkably, yielding 47.6\% compression for nodes, 49.7\% for edges and 45.7\% for core.

\subsection{Experiments on SAT Benchmarks}
\label{sec:compexp2}

A second stage of experimentation was preceded by an implementation of all the compression algorithms discussed so far (Alg.~\ref{alg:prop} - Alg.~\ref{alg:combalg3})
within a new tool, PeRIPLO~\cite{periplo}; PeRIPLO, built on MiniSAT 2.2.0, is an open-source SAT-solver which features resolution proof manipulation and interpolant generation capabilities
\footnote{Full experimental data, as well as executables used in tests are available at\\ \url{http://verify.inf.usi.ch/sites/default/files/RPTCI2014.tar.gz}}.

We evaluated the following algorithms: PushdownUnits (PU), RecyclePivotsWithIntersection (RPI), ReduceAndExpose (RE), StructuralHashing (SH) 
(Algs.~\ref{alg:simp},\ref{alg:recpiv},\ref{alg:rectra},\ref{alg:structhash}) and their combinations RPI+RE (Alg.~\ref{alg:combalg1}), 
SH+RE (Alg.~\ref{alg:combalg2}), PU+RPI+SH+RE (Alg.~\ref{alg:combalg3});
the evaluation was carried out on a set of purely propositional benchmarks
from the SAT Challenge 2012~\cite{SATCH12}, the SATLIB benchmark suite~\cite{SATLIB}
and the CMU collection~\cite{CMU}.

First, a subset of \emph{unsatisfiable} benchmarks was extracted from the SAT Challenge 2012 collection by running MiniSAT 2.2.0 alone with a 
timeout of 900 seconds and a memory threshold of 14GB;
this resulted in 261 instances from the Application track and the Hard Combinatorial track. 
In addition to these, another 125 unsatisfiable instances were obtained from the SATLIB Benchmark Suite and the CMU collection, for a total of 386 instances.

\begin{table}[!htp]
\caption{Results for SAT benchmarks. \#Bench reports the number 
         of benchmarks solved and processed within the time/memory constraints, RedNodes\% and
	 RedEdges\% report the average compression in the number of nodes and edges of the proof graphs,
         RedCore\% the average compression in the unsatisfiable core size. TranTime
         is the average transformation time in seconds; Ratio is the ratio between transformation time and overall time.}
         \medskip
\centering
\begin{tabular}{|p{1.7cm} || p{0.6cm} | p{0.6cm} ||  p{0.6cm} | p{0.6cm} || p{0.6cm} | p{0.6cm} || p{0.6cm} | p{0.6cm} | p{0.6cm} | p{0.6cm} | p{0.6cm} | p{0.6cm} |} 
\hline
& \multicolumn{2}{|c|}{\scriptsize{\#Bench}}
& \multicolumn{2}{|c|}{\scriptsize{RedNodes\%}}
& \multicolumn{2}{|c|}{\scriptsize{RedCore\%}}
& \multicolumn{2}{|c|}{\scriptsize{RedEdges\%}}
& \multicolumn{2}{|c|}{\scriptsize{TranTime(s)}}
& \multicolumn{2}{|c|}{\scriptsize{Ratio}} \\ 
\hline 
\hline 
\scriptsize{PU} & \multicolumn{2}{|c|}{200}
& \multicolumn{2}{|c|}{1.81}
& \multicolumn{2}{|c|}{0.00}
& \multicolumn{2}{|c|}{2.18}
& \multicolumn{2}{|c|}{5.44}
& \multicolumn{2}{|c|}{0.09}\\ 
\hline 
\hline 
\scriptsize{SH} & \multicolumn{2}{|c|}{205}
& \multicolumn{2}{|c|}{5.90}
& \multicolumn{2}{|c|}{0.00}
& \multicolumn{2}{|c|}{6.55}
& \multicolumn{2}{|c|}{4.53}
& \multicolumn{2}{|c|}{0.07}\\ 
\hline 
\hline 
\scriptsize{RPI} & \multicolumn{2}{|c|}{203}
& \multicolumn{2}{|c|}{28.48}
& \multicolumn{2}{|c|}{1.75}
& \multicolumn{2}{|c|}{30.66}
& \multicolumn{2}{|c|}{14.32}
& \multicolumn{2}{|c|}{0.21}\\ 
\hline 
\hline 
\scriptsize{RE} 3 & \multicolumn{2}{|c|}{203}
& \multicolumn{2}{|c|}{4.16}
& \multicolumn{2}{|c|}{0.09}
& \multicolumn{2}{|c|}{4.85}
& \multicolumn{2}{|c|}{24.84}
& \multicolumn{2}{|c|}{0.31}\\ 
\hline 
\hline 
\scriptsize{RE} 5 & \multicolumn{2}{|c|}{203}
& \multicolumn{2}{|c|}{5.06}
& \multicolumn{2}{|c|}{0.14}
& \multicolumn{2}{|c|}{5.88}
& \multicolumn{2}{|c|}{37.86}
& \multicolumn{2}{|c|}{0.41}\\ 
\hline 
\hline 
\scriptsize{RE} 10 & \multicolumn{2}{|c|}{202}
& \multicolumn{2}{|c|}{6.11}
& \multicolumn{2}{|c|}{0.17}
& \multicolumn{2}{|c|}{7.08}
& \multicolumn{2}{|c|}{67.09}
& \multicolumn{2}{|c|}{0.56}\\ 
\hline
\hline 
\scriptsize{PU+SH+RPI} & \multicolumn{2}{|c|}{196}
& \multicolumn{2}{|c|}{32.81}
& \multicolumn{2}{|c|}{1.47}
& \multicolumn{2}{|c|}{35.70}
& \multicolumn{2}{|c|}{18.66}
& \multicolumn{2}{|c|}{0.27}\\ 
\hline 
\end{tabular} 
\smallskip \\
(a)
\bigskip \\
\centering
\begin{tabular}{|p{1.2cm} || p{0.6cm} | p{0.6cm} ||  p{0.6cm} | p{0.6cm} || p{0.6cm} | p{0.6cm} || p{0.6cm} | p{0.6cm} | p{0.6cm} | p{0.6cm} | p{0.6cm} | p{0.6cm} |} 
\hline
\scriptsize{RPI+RE} & \multicolumn{2}{|c|}{\scriptsize{\#Bench}}
& \multicolumn{2}{|c|}{\scriptsize{RedNodes\%}}
& \multicolumn{2}{|c|}{\scriptsize{RedCore\%}}
& \multicolumn{2}{|c|}{\scriptsize{RedEdges\%}}
& \multicolumn{2}{|c|}{\scriptsize{TranTime(s)}}
& \multicolumn{2}{|c|}{\scriptsize{Ratio}} \\ 
\hline 
\hline 
2,3 & \multicolumn{2}{|c|}{201}
& \multicolumn{2}{|c|}{30.69}
& \multicolumn{2}{|c|}{2.08}
& \multicolumn{2}{|c|}{33.49}
& \multicolumn{2}{|c|}{34.78}
& \multicolumn{2}{|c|}{0.39}\\ 
\hline 
\hline 
2,5 & \multicolumn{2}{|c|}{200}
& \multicolumn{2}{|c|}{30.71}
& \multicolumn{2}{|c|}{2.15}
& \multicolumn{2}{|c|}{33.53}
& \multicolumn{2}{|c|}{40.37}
& \multicolumn{2}{|c|}{0.45}\\ 
\hline 
\hline 
3,3 & \multicolumn{2}{|c|}{200}
& \multicolumn{2}{|c|}{31.28}
& \multicolumn{2}{|c|}{2.23}
& \multicolumn{2}{|c|}{34.22}
& \multicolumn{2}{|c|}{51.43}
& \multicolumn{2}{|c|}{0.51}\\ 
\hline 
\hline 
3,5 & \multicolumn{2}{|c|}{200}
& \multicolumn{2}{|c|}{31.56}
& \multicolumn{2}{|c|}{2.34}
& \multicolumn{2}{|c|}{34.50}
& \multicolumn{2}{|c|}{61.16}
& \multicolumn{2}{|c|}{0.56}\\ 
\hline 
\end{tabular} 
\smallskip \\
(b)
\bigskip \\
\centering
\begin{tabular}{|p{1.2cm} || p{0.6cm} | p{0.6cm} ||  p{0.6cm} | p{0.6cm} || p{0.6cm} | p{0.6cm} || p{0.6cm} | p{0.6cm} | p{0.6cm} | p{0.6cm} | p{0.6cm} | p{0.6cm} |} 
\hline
\scriptsize{SH+RE} & \multicolumn{2}{|c|}{\scriptsize{\#Bench}}
& \multicolumn{2}{|c|}{\scriptsize{RedNodes\%}}
& \multicolumn{2}{|c|}{\scriptsize{RedCore\%}}
& \multicolumn{2}{|c|}{\scriptsize{RedEdges\%}}
& \multicolumn{2}{|c|}{\scriptsize{TranTime(s)}}
& \multicolumn{2}{|c|}{\scriptsize{Ratio}} \\ 
\hline 
\hline 
2,3 & \multicolumn{2}{|c|}{204}
& \multicolumn{2}{|c|}{17.33}
& \multicolumn{2}{|c|}{0.09}
& \multicolumn{2}{|c|}{19.20}
& \multicolumn{2}{|c|}{33.87}
& \multicolumn{2}{|c|}{0.38}\\ 
\hline 
\hline 
2,5 & \multicolumn{2}{|c|}{204}
& \multicolumn{2}{|c|}{19.81}
& \multicolumn{2}{|c|}{0.15}
& \multicolumn{2}{|c|}{21.92}
& \multicolumn{2}{|c|}{48.40}
& \multicolumn{2}{|c|}{0.47}\\
\hline 
\hline 
3,3 & \multicolumn{2}{|c|}{204}
& \multicolumn{2}{|c|}{21.68}
& \multicolumn{2}{|c|}{0.16}
& \multicolumn{2}{|c|}{23.96}
& \multicolumn{2}{|c|}{56.39}
& \multicolumn{2}{|c|}{0.51}\\ 
\hline 
\hline 
3,5 & \multicolumn{2}{|c|}{202}
& \multicolumn{2}{|c|}{23.69}
& \multicolumn{2}{|c|}{0.18}
& \multicolumn{2}{|c|}{26.17}
& \multicolumn{2}{|c|}{70.75}
& \multicolumn{2}{|c|}{0.59}\\
\hline 
\end{tabular} 
\smallskip \\
(c)
\bigskip \\
\centering
\begin{tabular}{|p{2.3cm} || p{0.6cm} | p{0.6cm} ||  p{0.6cm} | p{0.6cm} || p{0.6cm} | p{0.6cm} || p{0.6cm} | p{0.6cm} | p{0.6cm} | p{0.6cm} | p{0.6cm} | p{0.6cm} |} 
\hline
\scriptsize{PU+SH+RPI+RE} & \multicolumn{2}{|c|}{\scriptsize{\#Bench}}
& \multicolumn{2}{|c|}{\scriptsize{RedNodes\%}}
& \multicolumn{2}{|c|}{\scriptsize{RedCore\%}}
& \multicolumn{2}{|c|}{\scriptsize{RedEdges\%}}
& \multicolumn{2}{|c|}{\scriptsize{TranTime(s)}}
& \multicolumn{2}{|c|}{\scriptsize{Ratio}} \\ 
\hline 
\hline 
2,3 & \multicolumn{2}{|c|}{195}
& \multicolumn{2}{|c|}{39.46}
& \multicolumn{2}{|c|}{1.89}
& \multicolumn{2}{|c|}{43.34}
& \multicolumn{2}{|c|}{35.23}
& \multicolumn{2}{|c|}{0.44}\\ 
\hline 
\hline 
2,5 & \multicolumn{2}{|c|}{195}
& \multicolumn{2}{|c|}{40.46}
& \multicolumn{2}{|c|}{1.93}
& \multicolumn{2}{|c|}{44.49}
& \multicolumn{2}{|c|}{38.49}
& \multicolumn{2}{|c|}{0.46}\\ 
\hline 
\hline 
3,3 & \multicolumn{2}{|c|}{195}
& \multicolumn{2}{|c|}{41.68}
& \multicolumn{2}{|c|}{2.06}
& \multicolumn{2}{|c|}{45.86}
& \multicolumn{2}{|c|}{47.41}
& \multicolumn{2}{|c|}{0.51}\\ 
\hline 
\hline 
3,5 & \multicolumn{2}{|c|}{195}
& \multicolumn{2}{|c|}{42.41}
& \multicolumn{2}{|c|}{2.05}
& \multicolumn{2}{|c|}{46.71}
& \multicolumn{2}{|c|}{52.91}
& \multicolumn{2}{|c|}{0.54}\\ 
\hline 
\end{tabular} 
\smallskip \\
(d)
\label{tbl:resSAT} 
\end{table} 

The experiments were carried out on a 64-bit Ubuntu server featuring a Quad-Core
4GHz Xeon CPU and 16GB of memory; a timeout of 1200 seconds and a memory 
threshold of 14GB were put as limit to the executions.
The PeRIPLO framework was able to handle proofs up to 30 million nodes, as in the case of the \emph{rbcl\_xits\_07\_UNSAT} instance from the Application track
in the SAT Challenge 2012 collection.

Differently from the case of SMT benchmarks, we decided to specify as termination criterion an explicit amount of transformation traversals per global iteration,
focusing on the dependency between proofs size and time taken by the algorithms to move over proofs and compress them.

Table~\ref{tbl:resSAT} reports the performance of the compression techniques. Table~\ref{tbl:resSAT}a shows the results for the individual techniques PU, SH, RPI, RE, the latter tested for an 
increasing amount of transformation traversals (3, 5, 10), and the combination PU+SH+RPI without RE. Tables~\ref{tbl:resSAT}b,~\ref{tbl:resSAT}c,~\ref{tbl:resSAT}d 
respectively report on the combinations RPI+RE, SH+RE, PU+SH+RPI+RE:
in the first column, a pair $n,m$ indicates that $n$ global iterations and $m$ transformation traversals per global iteration were carried out.

RPI is clearly the most effective technique on a single run, as for compression and ratio transformation time / overall time. 
For this set of experiments we tuned RE focusing on its ability to disclose new redundancies, so we did not expect exceptional
results when running the algorithm by itself; the performance of RE improves with the number of transformation traversals performed, but cannot match that of RPI.

On the other hand, the heuristics adopted in the application of the rewriting rules (see \S\ref{sec:heuri}) have a major effect on SH, 
enhancing the amount of compression from about 6\%  to more than 20\%.

The combined approaches naturally achieve better and better results as the number of global iterations and transformation traversals grows.
In particular, Alg.~\ref{alg:combalg3}, which brings together the techniques for regularization, compactness and redundancies exposure, reaches a remarkable average compression level of 40\%, 
surpassing (ratio being equal) all other combined approaches. 

\begin{table}[!ht]
\caption{Results for SAT benchmarks. MaxRedNodes\% and MaxRedEdges\% are the maximum compression 
	 of nodes and edges achieved by the PU+SH+RPI+RE combination on a single benchmark.
	 }
	 \medskip
\centering
\begin{tabular}{|p{2.3cm} || p{0.6cm} | p{0.6cm} ||  p{0.6cm} | p{0.6cm} || p{0.6cm} | p{0.6cm} ||} 
\hline
\scriptsize{PU+SH+RPI+RE} & \multicolumn{2}{|c|}{\scriptsize{MaxRedNodes\%}}
& \multicolumn{2}{|c|}{\scriptsize{MaxRedCore\%}}
& \multicolumn{2}{|c|}{\scriptsize{MaxRedEdges\%}}\\
\hline 
\hline 
2,3 & \multicolumn{2}{|c|}{83.7}
& \multicolumn{2}{|c|}{21.5}
& \multicolumn{2}{|c|}{83.7}\\ 
\hline 
\hline 
2,5 & \multicolumn{2}{|c|}{84.9}
& \multicolumn{2}{|c|}{21.6}
& \multicolumn{2}{|c|}{85.2}\\ 
\hline 
\hline 
3,3 & \multicolumn{2}{|c|}{87.1}
& \multicolumn{2}{|c|}{22.1}
& \multicolumn{2}{|c|}{87.4}\\ 
\hline 
\hline 
3,5 & \multicolumn{2}{|c|}{87.9}
& \multicolumn{2}{|c|}{22.2}
& \multicolumn{2}{|c|}{88.2}\\ 
\hline 
\end{tabular}
\label{tbl:resSATallTOP}
\end{table}

We report for completeness in Table~\ref{tbl:resSATallTOP} the maximum compression obtained by the PU+SH+RPI+RE combination
on the best individual instances. 

A limitation of the current version of PeRIPLO it that preprocessing by SATElite is not enabled in case of proof-logging;
this restriction, which entails higher solving times and might yield larger proofs sizes, will be addressed in a future release of
the tool.

\section{Proof Transformation for Interpolation}
\label{sec:reordtrasf}

Craig interpolants \cite{Cra57}, since the seminal work by McMillan \cite{McM03,McM04a,McM04b}, have been extensively applied in SAT-based
model checking and predicate abstraction \cite{HJM+04}. Formally,
given an unsatisfiable conjunction of formulae $A \wedge B$, an interpolant $I$ is a formula that is implied by $A$ (i.e., $A \implies I$), 
is unsatisfiable in conjunction with $B$ (i.e., $B \wedge I \implies \bot$) and
is defined on the common language of $A$ and $B$. 
The interpolant $I$ can be thought of as an over-approximation of
$A$ that is still in conflict with $B$.

Several state-of-the art approaches exist to generate interpolants in an automated manner;
the most successful techniques derive an interpolant for $A \wedge B$ from a proof of unsatisfiability of the conjunction.  
This approach grants two important benefits: the generation can be achieved in linear time w.r.t. the proof size, and interpolants themselves
only contain information relevant to determine the unsatisfiability of $A \wedge B$.

Pudl\'ak and Kraj\'{i}\v{c}ek~\cite{Pud97,Kra97} are 
probably the first to propose an efficient way to compute interpolants in the
context of propositional logic. McMillan~\cite{McM04b} proposes an alternative
method that also handles the quantifier-free theories of uninterpreted 
functions, linear arithmetic, and their combination.
All these techniques adopt recursive algorithms, which initially
set \emph{partial interpolants} for the axioms.
Then, following the proof structure, they deduce a partial interpolant for each conclusion from those of the premises. 
The partial interpolant of the overall conclusion is the interpolant for the formula.

Yorsh and Musuvathi present in~\cite{YM05} a generalization of Pudl\'ak's
method that can compute interpolants for a formula defined modulo a 
theory \T. The leaves of the proof of unsatisfiability in this case
are original clauses as well as \T-lemmata involving original predicates, generated 
by the prover during the solving process. It is then sufficient to compute a 
partial interpolant for each theory lemma in order to derive the global 
interpolant.

The last technique, for its modularity, finds its natural implementation 
within SMT-solvers~\cite{BSS+09}, procedures that combine SAT-solvers and 
domain specific algorithms for a theory \T in an efficient way (see \S\ref{sec:smt}). 
Cimatti et al.~\cite{CGS08} show that interpolant generation within
SMT-solvers can outperform other known methods (e.g. \cite{McM04b}),
as a result of using optimized domain-specific procedures for \T.

In the following we use $A$ and $B$ to denote two quantifier-free
formulae in a theory \T, for which we would like to compute an
interpolant. Theories of interest are equality with uninterpreted functions
\EUF, linear arithmetic over the rationals \LRA and the integers \LIA,
the theory of arrays \AX, or a combination of theories, such as 
$\EUF \cup \LRA$. Variables that appear only in $A$ or $B$ are called 
$A$-\emph{local} and $B$-\emph{local} respectively. Variables that appear in both
$A$ and $B$ are called $AB$-\emph{common}. A predicate is called $AB$-\emph{mixed} if it 
is defined on both $A$-local and $B$-local variables, it is called $AB$-\emph{pure} 
otherwise. Notice that $AB$-mixed predicates cannot appear in $A$ and $B$.
\begin{example}
  Let $A \equiv (x=v \wedge f(x)=z)$, 
  $B \equiv (y=v \wedge f(y)=u \wedge z\not=u)$ be two formulae
  in the \EUF theory.
  Variable $x$ is $A$-local, $y,u$ are $B$-local, $z,v$ are $AB$-common
  (a predicate $x=y$ would be $AB$-mixed).
  An interpolant $I$ for $A\wedge B$ is $f(v)=z$, which is an $AB$-pure
  predicate.
\label{ex:euf}
\end{example}
We consider resolution proofs are defined as in \S\ref{sec:smt};
recall that propositional variables in a
proof may represent the propositional abstraction of theory 
predicates. In this case we say
that a propositional variable is $AB$-mixed if such is the predicate associated with it.

One limitation of the approach of~\cite{YM05} is that theory lemmata,
appearing in a proof of unsatisfiability, must not contain $AB$-mixed
predicates. However, several decision procedures defined for SMT-solvers
heavily rely on the creation of new predicates during the solving process.
Examples are delayed theory combination (DTC)~\cite{BBC+05c}, Ackermann's
Expansion~\cite{Ack54}, Lemmas on Demand~\cite{dMR02} and Splitting on 
Demand~\cite{BNO+06} (see \S\ref{subsec:applications}). All these methods 
may introduce new predicates, which can potentially be  
$AB$-mixed.

In this section we show how to compute an $AB$-pure proof 
from an $AB$-mixed one but without interfering with the internals of
the SMT-solver; our technique applies to any approach that requires the addition of
$AB$-mixed predicates (see \S\ref{subsec:applications} for a set of
examples). We illustrate how to employ the Local Transformation Framework to
effectively modify the proofs, in such a way that the generic method
of~\cite{YM05} can be applied; in this way it is possible to achieve 
a complete decoupling between the solving phase and the interpolant generation 
phase, provided that an interpolation procedure is
available for a conjunction of atoms in $\T$.

\begin{figure}
\centering
\scalebox{.45}{\input{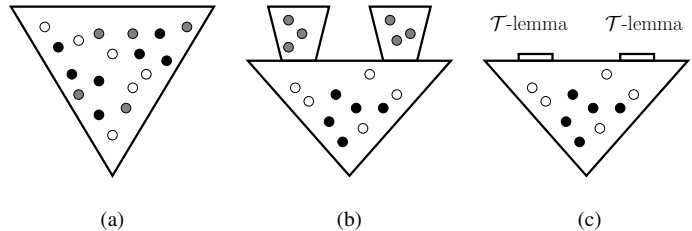}}
\caption{An overview of our approach. (a) is the proof
generated by the SMT-solver. White points represent $A$-local
predicates, black points represent $B$-local predicates, 
grey points represent $AB$-mixed predicates. (b) $AB$-mixed
predicates are confined inside $AB$-mixed trees. (c)
$AB$-mixed trees are removed and their roots are valid
theory lemmata in \T.}
\label{fig:overview}
\end{figure}

A sketch of the approach is depicted in Fig.~\ref{fig:overview}.
The idea is to move all $AB$-mixed predicates (in grey) toward the leaves
of the proof (Figure~\ref{fig:overview}b) within maximal $AB$-mixed subproofs. 

\begin{definition}[$AB$-mixed subproof]
\label{def:maximal}
Given a resolution proof $P$, an {\em $AB$-mixed subproof} is a 
subproof $P'$ of $P$ rooted in a clause $C$, whose intermediate pivots 
are all $AB$-mixed predicates. $P'$ is {\em maximal} if $C$ does not contain $AB$-mixed predicates.
\end{definition}

When dealing with a background theory \T we note the following fact:
if $P'$ is a maximal $AB$-mixed subproof rooted in a clause $C$, 
then \emph{$C$ is a valid theory lemma for \T}.

This observation derives from Def.~\ref{def:maximal} and from 
the fact that ($i$) $AB$-mixed predicates can only appear in theory 
lemmata (as they do not appear in the original formula) and 
($ii$) a resolution step over two theory lemmata generates 
another theory lemma.

Once $AB$-mixed maximal subproofs are formed, it is possible to 
replace them with their root clauses (Figure~\ref{fig:overview}c).
The obtained proof is now free of $AB$-mixed predicates and can
be used to derive an interpolant applying the method of~\cite{YM05},
provided that an interpolant generating procedure is available for the theory \T.

The crucial part of our approach is an algorithm for
proof transformation. It relies on the Local Transformation Framework
discussed in~\S\ref{sec:tra_fra}.
An ad-hoc application of the rules can be used to transform 
a proof $P$ into a proof $P'$, where all $AB$-mixed variables are confined in
$AB$-mixed subproofs. Each rewriting rule can effectively swap two pivots
$p$ and $q$ in the resolution proof, or perform simplifications, 
depending on the particular context. 

In the following, to facilitate the
understanding of the algorithm, we will call $AB$-mixed and $AB$-pure
predicates {\em light} and {\em heavy} respectively. The rules
are applied when a light predicate is below a heavy predicate in
the proof graph. The effect of an exhaustive application of the rules 
is to lift light predicates over heavy predicates as bubbles in water.

\subsection{Pivot Reordering Algorithms}
\label{subsec:pivreo}

The Local Transformation Framework can be effectively employed to perform a \emph{local reordering} 
of the pivots. Each rule in Fig.~\ref{fig:rulesdag} either swaps the position of two pivots ($S1$, $S2$, $R2$), 
or it eliminates at least one pivot ($R1$, $R2'$, $R3$).
This feature can be used to 
create an application strategy aimed at sorting the pivots in a proof $P$,
by transforming it into a proof
$P'$ such that all light variables are moved above
heavy variables.

In order to achieve this goal it is sufficient to consider
only {\em unordered} contexts, i.e. those in which $v(t)$ 
is a light variable and $v(s)$ is a heavy variable. 
Therefore a simple non-deterministic algorithm can be 
derived as Alg.~\ref{alg:reor}.

\begin{algorithm}[!ht]
\SetAlgoNoLine
\LinesNumbered
\DontPrintSemicolon
\KwIn{A legal proof}
\KwOut{A legal proof without unordered contexts}
\KwData{$U$: set of unordered contexts}
\Begin{
Determine $U$, e.g. through a visit of the proof\;
\While{$U \neq \emptyset$}{
Choose a context in $U$\;
Apply the associated rule, and SubsumptionPropagation if necessary\;
Update $U$\;
}
}
\caption{PivotReordering.}
\label{alg:reor}
\end{algorithm}

The algorithm terminates: note in fact that each iteration 
strictly decreases the distance of an occurrence of a heavy 
pivot w.r.t. the global root, until no more unordered contexts are left.

\begin{algorithm}[!ht]
\SetAlgoNoLine
\LinesNumbered
\DontPrintSemicolon
\KwIn{A left context $lc$, a right context $rc$}
\Begin{
\uIf{$lc$ is ordered and $rc$ is unordered}
{Apply rule for $rc$\;}
\uElseIf{$lc$ is unordered and $rc$ is ordered}
{Apply rule for $lc$\;}
\uElseIf{$lc$ is unordered and $rc$ is unordered}
{Heuristically choose between $lc$ and $rc$ and apply rule\;}
}
\caption{ApplyRuleForPivotReordering.}
\label{alg:aprul-reor}
\end{algorithm}

A more efficient choice is to make use of Alg.~\ref{alg:rectraloop} TransformAndReconstruct,
by instantiating the ApplyRule method so that it systematically pushes light variables above heavy ones; 
a possible instantiation is shown in Alg.~\ref{alg:aprul-reor}.
An algorithm for pivot reordering would then consist of a number of consecutive runs of TransformAndReconstruct,
stopping when no more unordered contexts are found: Alg.~\ref{alg:reor2}, \emph{PivotReordering2}, implements this approach.

\begin{algorithm}[!ht]
\SetAlgoNoLine
\LinesNumbered
\DontPrintSemicolon
\KwIn{A legal proof}
\KwOut{A legal proof without unordered contexts}
\Begin{
\While{unordered contexts are found}{
TransformAndReconstruct($ApplyRuleForPivotReordering$)\;
}
}
\caption{PivotReordering2.}
\label{alg:reor2}
\end{algorithm}

\subsection{SMT-Solving and $AB$-Mixed Predicates}
\label{subsec:applications}

In this section we show a number of techniques currently
employed in state-of-the-art SMT-solvers that can potentially
introduce $AB$-mixed predicates during the solving phase. If 
these predicates become part of the proof of unsatisfiability, 
the proof reordering algorithms described in \S\ref{subsec:pivreo} can be
applied to produce an $AB$-pure proof. 

\subsubsection{Theory Reduction Techniques}

Let $\T_k$ and $\T_j$ be two 
decidable theories such that 
$\T_k$ is weaker (less expressive) than $\T_j$. 
Given a $\T_j$-formula $\varphi$, and a decision procedure $\mbox{SMT}(\T_k)$ for 
quantifier-free formulae in $\T_k$, 
it is often possible to obtain a decision procedure $\mbox{SMT}(\T_j)$ for quantifier-free
formulae in $\T_j$ by augmenting 
$\varphi$ with a finite set of $\T_k$-lemma $\psi$. These lemmata (or axioms) 
explicitly encode the necessary knowledge such that $\T_k \models \varphi \wedge \psi$
if and only if $\T_j \models \varphi$. Therefore a simple decision procedure for $\T_j$
is as described by Alg.~\ref{alg:SMTT2}.

\begin{algorithm}[!ht]
\SetAlgoNoLine
\LinesNumbered
\DontPrintSemicolon
\KwIn{$\varphi$ for $\T_j$}
\Begin{
  $\psi$ = generateLemmata($\varphi$)\;
  \KwRet{$\mbox{SMT}(\T_k)( \varphi \wedge \psi )$}\;
}
\caption{A reduction approach for SMT($\T_j$).}
\label{alg:SMTT2}
\end{algorithm}

In practice the lemmata generation function can be made
lazy by plugging it inside the SMT-solver directly; this
paradigm is known as Lemma on Demand~\cite{dMR02} or
Splitting on Demand~\cite{BNO+06}. 
We show some reduction techniques as follows.

\begin{figure*}[!htp]
  \centering
  \begin{tabular}{|l|l|l|}
    \hline
    Id & Clauses                           & Prop. abstract. \\
    \hline
    1 & $x=wr(y,i,e)$                      & ${p}_1$          \\
    2 & $rd(x,j) \neq rd(y,j)$            & $\n{{p}_2}$ \\
    3 & $rd(x,k) \neq rd(y,k)$            & $\n{{p}_3}$ \\
    4 & $j\neq k$                          & $\n{{p}_4}$ \\
    5 & $(i=j \vee rd(wr(y,i,e),j)=rd(y,j))$ & ${p}_5\, {p}_6$ \\
    6 & $(i=k \vee rd(wr(y,i,e),k)=rd(y,k))$ & ${p}_7\, {p}_8$ \\
    7 & $(x\neq wr(y,i,e) \vee rd(x,j)= rd(y,j) \vee rd(wr(y,i,e),j)\neq rd(y,j))$ & $\n{{p}_1}\, {p}_2\, \n{{p}_6}$ \\
    8 & $(x\neq wr(y,i,e) \vee rd(x,k)= rd(y,k) \vee rd(wr(y,i,e),k)\neq rd(y,k))$ & $\n{{p}_1}\, {p}_3\, \n{{p}_8}$ \\
    9 & $(j=k \vee i\neq j \vee i\neq k)$                                   & ${p}_4\, \n{{p}_5}\, \n{{p}_7}$ \\
    \hline
  \end{tabular}
  \bigskip \\
  \begin{tabular}{c}
  \begin{minipage}{0.8\textwidth}
  \scriptsize
  \def\defaultHypSeparation{\hskip .1in}
  \begin{prooftree}
    \AxiomC{$p_7 p_8$}
					 \AxiomC{$\n{p_1} p_3 \n{p_8}$} 
    \BinaryInfC{$\n{p_1} p_3 p_7$}
                                         \AxiomC{$p_4 \n{p_5} \n{p_7}$}
    \BinaryInfC{$\mathbf{\n{p_1} p_3 p_4 \n{p_5}}$}
    \LeftLabel{$\mathbf{p_4}$}
                                         \AxiomC{$\mathbf{\n{p_4}}$}
    \BinaryInfC{$\mathbf{\n{p_1} p_3 \n{p_5}}$}
										  \AxiomC{$p_5 p_6$}
												    \AxiomC{$\n{p_1} p_2 \n{p_6}$}
										  \BinaryInfC{$\mathbf{\n{p_1} p_2 p_5}$}
    \LeftLabel{$\mathbf{p_5}$}
    \BinaryInfC{$\mathbf{\n{p_1} p_2 p_3}$}
					\AxiomC{$\n{p_3}$}
    \BinaryInfC{$\n{p_1} p_2$}
					\AxiomC{$p_1$}
    \BinaryInfC{$p_2$}
					\AxiomC{$\n{p_2}$}
    \BinaryInfC{$\bot$}
  \end{prooftree}
  \end{minipage}
\\   \\  (a) \\ \\ \\ \\
  \begin{minipage}{0.8\textwidth}
  \scriptsize
    \def\defaultHypSeparation{\hskip .1in}
  \begin{prooftree}
    \AxiomC{$p_7 p_8$}
					 \AxiomC{$\n{p_1} p_3 \n{p_8}$} 
    \BinaryInfC{$\n{p_1} p_3 p_7$}
                                         \AxiomC{$p_4 \n{p_5} \n{p_7}$}
    \BinaryInfC{$\mathbf{\n{p_1} p_3 p_4 \n{p_5}}$}
										  \AxiomC{$p_5 p_6$}
												    \AxiomC{$\n{p_1} p_2 \n{p_6}$}
										  \BinaryInfC{$\mathbf{\n{p_1} p_2 p_5}$}
    \LeftLabel{$\mathbf{p_5}$}
    \BinaryInfC{$\mathbf{\n{p_1} p_2 p_3 p_4}$}
    \LeftLabel{$\mathbf{p_4}$}
                                         \AxiomC{$\mathbf{\n{p_4}}$}
    \BinaryInfC{$\mathbf{\n{p_1} p_2 p_3}$}
					\AxiomC{$\n{p_3}$}
    \BinaryInfC{$\n{p_1} p_2$}
					\AxiomC{$p_1$}
    \BinaryInfC{$p_2$}
					\AxiomC{$\n{p_2}$}
    \BinaryInfC{$\bot$}
  \end{prooftree}
  \end{minipage}
  \\ \\
(b)
  \end{tabular}

  \caption{Clauses from Ex.~\ref{ex:axexample}. $\varphi \equiv \{ 1,2,3,4 \}$, $\psi \equiv \{ 5,6 \}$. Clauses 7-9 are
           theory lemmata discovered by the $\EUF$ solver. (a) is a possible proof obtained by the SMT-solver (for \EUF) 
	   on $\varphi \wedge \psi$. (b) is a proof after swapping $p_4$ and $p_5$ by means of rule $S2$; in the
	   resulting proof all mixed literals ($p_5$-$p_8$) appear in the upper part of the proof in an $AB$-mixed
	   proof subtree. The root of the $AB$-mixed subtree $\n{p_1} p_2 p_3 p_4$ is a valid theory 
	   lemma in \AX.} 
  \label{fig:axproof}
\end{figure*}

\paragraph{Reduction of \AX to \EUF.}

We consider the case where $\T_k \equiv \EUF$, the theory
of equality with uninterpreted functions, and $\T_j \equiv \AX$, 
the theory of arrays with extensionality. The axioms of \EUF
are the ones of equality (reflexivity, symmetry, and transitivity)
plus the congruence axioms $\forall x, y.\ x=y \implies f(x)=f(y)$, 
for any functional symbol of the language.

The theory of arrays \AX is instead axiomatized by:
\begin{eqnarray}
  \label{ax1}
  \forall x, i, e. & & rd(wr(x,i,e),i)=e \\
  \label{ax2}
  \forall x, i, j, e. & & i = j \vee rd(wr(x,i,e),j)=rd(x,j) \\
  \label{ax3}
  \forall x, y. & & x=y \iff (\forall i.\ rd(x,i)=rd(y,i))
\end{eqnarray}

State-of-the-art approaches for \AX implemented
in SMT-solvers~\cite{BB08,dMB09,GKF08,BNO+08b}
are all based on reduction to \EUF. Instances
of the axioms of \AX are added to the formula
in a lazy manner until either the formula is proven
unsatisfiable or saturation is reached. The addition
of new lemmata may require the creation of
$AB$-mixed predicates when a partitioned formula
is considered.

\begin{example}
\label{ex:axexample}
  Let $\varphi \equiv A \wedge B$, where $A \equiv x=wr(y,i,e)$, and 
  $B \equiv rd(x,j) \not= rd(y,j) \wedge rd(x,k) \not= rd(y,k) \wedge j\not=k$.
  Variables $\{ i, e \}$ are $A$-local, $\{ j,k \}$ are $B$-local, and $\{ x,y \}$ are
  $AB$-common. To prove $\varphi$ unsatisfiable with a reduction
  to \EUF, we need to instantiate axiom~(\ref{ax2}) twice as
  $\psi \equiv (i=j \vee rd(wr(y,i,e),j)=rd(y,j)) \wedge (i=k \vee rd(wr(y,i,e),k)=rd(y,k))$. 
  Notice that we introduced four $AB$-mixed predicates. Now we can send 
  $\varphi \wedge \psi$ to an SMT-solver for \EUF to produce the proof of unsatisfiability. 
  Fig.~\ref{fig:axproof} shows a possible resolution proof generated by the SMT-solver,
  and how it can be transformed into a proof without $AB$-mixed predicates.
\end{example}

\paragraph{Reduction of \LIA to \LRA.}

Decision procedures for \LIA (linear integer arithmetic)
often rely on iterated calls to a decision procedure for
\LRA (linear rational arithmetic). An example is the 
method of \emph{branch-and-bound}: given a feasible rational
region $R$ for a set of variables $\vec{x} = (x_1,\ldots,x_n)$, 
and a non-integer point $\vec{c} \in R$ for $\vec{x}$, then one
step of branch-and-bound generates the two subproblems
$R \cup \{ x_i \leq \lfloor c_i \rfloor \}$ and 
$R \cup \{ x_i \geq \lceil c_i \rceil \}$. These 
are again recursively explored until an integer point $\vec{c}$
is found. 

Note that the splitting on the bounds can be delegated to
the propositional engine by adding the lemma 
$((x_i \leq \lfloor c_i \rfloor) \vee (x_i \geq \lceil c_i \rceil))$.
In order to obtain a faster convergence of the algorithm, it
is possible to split on {\em cuts}, i.e. linear constraints, rather 
than on simple bounds. However cuts may add $AB$-mixed predicates
if $A$-local and $B$-local variables are mixed into the same cut.

\begin{example}
Let $\varphi \equiv A \wedge B$ in \LIA, where 
$A \equiv 5x - y \leq 1 \wedge y - 5x \leq -1$, and 
$B \equiv 5z - y \leq -2 \wedge y - 5z \leq 3$. 
The axiom $\psi \equiv ((x-z \leq 0) \vee (x-z \geq 1))$
(which contains two $AB$-mixed literals)
is sufficient for $\varphi \wedge \psi$ to be
proven unsatisfiable by a solver for \LRA, by
discovering two additional theory lemmata
$((5x - y \not\leq 1) \vee 
  (y - 5z \not\leq 3) \vee
  (x-z \leq 0))$
  and
$((5x - y \not\leq -1) \vee 
  (y - 5z \not\leq -2) \vee
  (x-z \geq 1))$.
\end{example}

\paragraph{Ackermann's Expansion.}

When $\T_j$ is a combination of theories of the form $\EUF \cup \T_k$,
Ackermann's expansion~\cite{Ack54} can be used to reduce the reasoning
from $\T_j$ to $\T_k$. The idea is to use as $\psi$ the exhaustive
instantiation of the congruence axiom $\forall x,y \ (x=y \implies f(x)=f(y))$ for all
pairs of variables appearing in uninterpreted functional symbols and 
all uninterpreted functional symbols $f$ in $\varphi$. This instantiation 
generates $AB$-mixed predicates when $x$ is instantiated with an $A$-local symbol and $y$ with a $B$-local one.

\begin{example}
  \label{ex:ack}
  Let $\T_k \equiv \LRA$.
  Let $\varphi \equiv A \wedge B$ and 
  $A \equiv (a=x+y \wedge f(a)=c)$, 
  $B \equiv (b=x+y \wedge f(b)=d \wedge c\not=d)$.
  The axiom $\psi \equiv ((a\not=b) \vee (f(a)=f(b))$ 
  is sufficient for \LRA to detect the unsatisfiability
  of $\varphi \wedge \psi$, by discovering two
  additional theory lemmata
  $((f(a)\not=f(b)) \vee (f(a)\not=c) \vee (f(b)\not=d) \vee (c\not=d))$
  and
  $((a\not=x+y) \vee (b\not=x+y) \vee (a=b))$.
\end{example}

\subsubsection{Theory Combination via DTC}

A generic framework for theory combination was introduced
by Nelson and Oppen in~\cite{NO79}. We recall it briefly as
follows.

Given two signature-disjoint
and stably-infinite theories $\T_1$ and $\T_2$, a decision procedure 
for a conjunction of constraints in the combined theory $\T_1 \cup \T_2$ 
can be obtained from the decision procedures for $\T_1$ and $\T_2$.
First, the formula $\varphi$ is {\em flattened}, i.e. auxiliary variables
are introduced to separate terms that contain both symbols of $\T_1$ and $\T_2$.
Then the idea is that the two theory solvers for $\T_1$ and $\T_2$ are forced to 
exhaustively exchange 
{\em interface equalities} i.e. equalities 
between {\em interface variables} (interface variables are those 
that appear both in constraints of $\T_1$ and $\T_2$ after 
flattening)\footnote{Note that in practice flattening can be avoided. For instance
in Ex.~\ref{ex:dtc} we do not perform any flattening.}. 

Delayed Theory Combination ($DTC$) implements a non-deterministic version
of the Nelson-Oppen framework, in which interface equalities are not
exchanged by the deciders directly, but they are guessed by the SAT-solver.
With DTC it is possible to achieve a higher level of modularity w.r.t.
the classical Nelson-Oppen framework. DTC is currently implemented (with
some variations) in most state-of-the-art SMT-solvers.

If no $AB$-mixed interface equality is generated, an interpolant
can be derived with the methods already present in the literature; otherwise our method
can be applied to reorder the proof, as an alternative to the techniques described
in~\cite{CGS08,GKT09}.

\begin{example}
\label{ex:dtc}
Consider again $\varphi$ of Ex.~\ref{ex:ack}. 
Since $a$, $b$, $f(a)$, $f(b)$ appear in
constraints of both theories, we need to generate
two interface equalities $a=b$ and $f(a)=f(b)$. The
guessing of their polarity is delegated to the SAT-solver.
The SMT-solver will detect the unsatisfiability after the
\EUF-solver discovers the two theory lemmata $((a\not=b) \vee (f(a)=f(b))$ 
and $((f(a)\not=f(b)) \vee (f(a)\not=c) \vee (f(b)\not=d) \vee (c\not=d))$
and the \LRA-solver discovers the theory lemma
$((a\not=x+y) \vee (b\not=x+y) \vee (a=b))$.
\end{example}

\subsection{Experiments on SMT benchmarks}
\label{subsec:reordexperiments}

For the purpose of this experimentation we chose to focus on one
particular application among those of \S\ref{subsec:applications}, namely
Ackermann's Expansion for Theory Combination.

We evaluated the proof transformation technique on the set of
\UFIDL formulae from the SMT-LIB~\cite{SMTLIB} 
(\UFIDL refers to the combined theory $\EUF \cup \IDL$).
The suite contains 319 unsatisfiable instances.
Each formula was split in half to obtain
an artificial interpolation problem (in the same fashion 
as~\cite{CGS08})\footnote{\label{ftn:website} The benchmarks and the
detailed results are available at \url{http://verify.inf.usi.ch/sites/default/files/RPTCI2014.tar.gz}}.

The pivot reordering algorithm Alg.~\ref{alg:reor2} was realized by means of the Local Transformation Framework  and implemented in \opensmt~\cite{BPS+10}.
Proof manipulation was applied when the proof contained $AB$-mixed predicates, in order to 
lift them up inside $AB$-maximal subproofs and replace them with their roots.


We ran the experiments on a 32-bit Ubuntu server equipped with Dual-Core
2GHz Opteron 2212 CPU and 4GB of memory. The benchmarks were executed 
with a timeout of 60 minutes and a memory threshold of
2GB (whatever was reached first): 172 instances, of which 82 proofs contained $AB$-mixed
predicates\footnote{Notice that in some cases $AB$-mixed predicates were produced
during the search, but they did not appear in the proof.}, were successfully handled
within these limits. We have reported the cost of the transformation and its effect 
on the proofs; the results are summarized in Table~\ref{tbl:ufidl_stat}. We
grouped benchmarks together following the original classification used in SMT-LIB
and provided average values for each group\footref{ftn:website}.
\begin{table}
\centering
\caption{The effect of proof transformation on QF\_UFIDL benchmarks summarized 
         per group: \#Bench - number of benchmarks in a group, \#$AB$ - average 
	 number of $AB$-mixed predicates in a proof, Time\% - average 
	 time overhead induced by  transformation, Nodes\% and 
	 Edges\% - average difference in the proof size as a result of transformation.}
	 \medskip
\begin{tabular}{|p{0.2\textwidth}|c|c|c|c|c|}

\hline
Group & \scriptsize{\#Bench} & \scriptsize{\#$AB$} & \scriptsize{Time\%} &  \scriptsize{Nodes\%} & \scriptsize{Edges\%}  \\
\hline\hline
\scriptsize{RDS} &      2 &       7 &      84 &     -16 &     -19 \\ \hline
\scriptsize{EufLaArithmetic} &      2 &      74 &      18 &     187 &     193 \\ \hline
\scriptsize{pete} &     15 &      20 &      16 &      66 &      68 \\ \hline
\scriptsize{pete2} &     52 &      13 &       6 &      73 &      80 \\ \hline
\scriptsize{uclid} &     11 &      12 &      29 &      87 &      90 \\ \hline
\hline
\scriptsize{Overall} &     82 &      16 &      13 &      74 &      79 \\ \hline
\end{tabular}
	 \label{tbl:ufidl_stat}
\end{table}

The results in Table~\ref{tbl:ufidl_stat} demonstrate that our proof
transformation technique induces, on average, about 13\% overhead with respect to
plain solving time.
The average increase in size is around 74\%, but not all the instances experienced 
a growth;  we observed in fact that in 42 out of 82 benchmarks the transformed 
proof was smaller than the original one both in the number of nodes and edges. Overall 
it is important to point out that the creation of new nodes due to the application of the 
$S$ rules did not entail any exponential blow-up
in the size of the proofs during the transformation process.

Another interesting result to report is the fact that only 45\% of the proofs 
contained $AB$-mixed predicates and, consequently, 
required transformation. This is another motivation
for using off-the-shelf algorithms for SMT-solvers and have the proof transformed 
in a second stage, rather than tweaking (and potentially slowing down) the solver 
to generate clean proofs upfront.

\subsection{Pivot Reordering for Propositional Interpolation}
This section concludes our discussion on interpolation by moving back from the context of SMT to that of SAT. 
We complete the analysis begun by the authors of~\cite{JM05} and
illustrate how, in the case of purely propositional refutations,
a transformation technique can be devised to  generate interpolants directly in conjunctive or disjunctive normal form.


Assuming a refutation of a formula $A \wedge B$, we distinguish whether a variable $p$ is local to $A$ ($p\in A$), local to $B$ ($p\in B$)
or common to $A$ and $B$ ($p\in AB$).
Fig.~\ref{fig:mcm} shows McMillan interpolation algorithm for propositional logic~\cite{McM04b,DKPW10}. 
The algorithm initially sets a partial interpolant for the clauses that label the refutation leaves; in particular, 
the partial interpolant of a clause in $A$ is its restriction $C |_{AB}$ to the propositional variables common to $A$ and  $B$. 
Then, recursively, a partial interpolant for each resolvent is computed from those of the antecedents depending on whether the pivot
appears only in $A$ ($I_1 \vee I_2$) or not ($I_1 \wedge I_2$); 
the partial interpolant of the global root is the interpolant for $A \wedge B$.
\begin{figure}[!ht]
\centering
\begin{minipage}[t]{.33\textwidth}
\begin{tabular}{|l c|c|c|c|}%
\hline
Leaf: & \multicolumn{4}{c|}{$C \, [I]$} \\
\hline 
\multicolumn{5}{|c|}{$I =
\left\{
	\begin{array}{ll}
		C|_{AB}  & \quad \mbox{if } C \in A\\
		\top & \quad \mbox{if } C \in B\\
	\end{array}
\right. $
}\\
\hline
\end{tabular}
\end{minipage}
\qquad
\begin{minipage}[t]{.60\textwidth}
\begin{tabular}{|l c|c|c|c|}%
\hline
Inner node: & \multicolumn{4}{c|}{$\quad \dfrac{C_1 \vee p \, [I_1] \qquad  C_2 \vee \n{p} \, [I_2]}{C_1 \vee C_2 \, [I]}$} \\
\hline
\multicolumn{5}{|c|}{$I =
\left\{
	\begin{array}{ll}
		I_1 \vee I_2  & \quad \mbox{if } p \in A\\
		I_1 \wedge I_2 & \quad \mbox{if } p \in B \text{ or } p \in AB
	\end{array}
\right. $
}\\
\hline
\end{tabular}
\end{minipage}
\caption{McMillan interpolation algorithm.}
\label{fig:mcm}
\end{figure}
In Fig.~\ref{fig:mcm}, $C[I]$ means that clause $C$ has a partial interpolant $I$.
$I_1$, $I_2$ and $I$ are the partial interpolants respectively associated with
the two antecedents $C_1 \vee p$, $C_2 \vee \n{p}$ and the resolvent $C_1 \vee C_2$ of a resolution step. 

Alg.~\ref{alg:reor2}, PivotReordering2, can be employed to restructure a refutation so that  McMillan interpolation algorithm generates an interpolant in CNF.
It is sufficient in fact to modify the definition of light and heavy predicates given in~\S\ref{sec:reordtrasf}, 
so that a context is considered unordered whenever $v(t)$ is local to $A$ (\emph{light}) and $v(s)$ is a
propositional variable in $B$ or in $AB$ (\emph{heavy}). Effect of the proof transformation is to push up light variables, so that, along every path from the leaves to the root,
light variables appear before heavy variables.

We need to show that this condition is sufficient in order for McMillan algorithm to produce an interpolant in CNF.
\begin{thm}
\label{thm:cnfization}
Assume a refutation $P$ without unordered contexts.  McMillan interpolation algorithm generates an interpolant in CNF from $P$.
\begin{proof}[by structural induction]
\paragraph{\textbf{Base case}.} The partial interpolant for a leaf labeled by a clause $C$ is either $\top$ or $C|_{AB}$, so it is in CNF.

\paragraph{\textbf{Inductive step}.} Given an inner node $n$ and the associated pivot $p=piv(n)$, assume the partial interpolants $I_1$ and $I_2$ for $C(n^+)=C_1 \vee p$ 
and $C(n^-)=C_2 \vee \n{p}$ are in CNF.
We have four possibilities:
\begin{itemize}
 \item Case 1: $I_1$ and $I_2$ are both in clausal form; then either $n^+,n^-$ are leaves or they are inner nodes with light pivot variables. $p$ can be either light or heavy:
 in the first case $I$ is itself a clause, in the second case $I$ is a conjunction of clauses, so it is in CNF.
 \item Case 2: $I_1$ is a clause, $I_2$ is a conjunction of at least two clauses; then $n^+$ can be either a leaf or an inner node with a light pivot, but $I_2$ must be an inner node with
 a heavy pivot (due to $\wedge$ being the main connective of $I_2$). Since $P$ does not have unordered contexts, $p$ must be a heavy variable, thus $I=I_1 \wedge I_2$ is in CNF.
 \item Case 3: $I_1$ is a conjunction of at least two clauses, $I_2$ is a clause. Symmetric to Case 2.
 \item Case 4: Both $I_1$ and $I_2$ are a conjunction of at least two clauses. As for Case 2 and Case 3.
\end{itemize}
\end{proof}
\end{thm}
A similar argumentation holds for the generation of interpolants in disjunctive normal form.
Let us consider the algorithm dual to McMillan, which we address as McMillan$'$~\cite{DKPW10}, illustrated in Fig.~\ref{fig:mcmp}.
\begin{figure}[!ht]
\centering
\begin{minipage}[t]{.33\textwidth}
\begin{tabular}{|l c|c|c|c|}%
\hline
Leaf: & \multicolumn{4}{c|}{$C \, [I]$} \\
\hline 
\multicolumn{5}{|c|}{$I =
\left\{
	\begin{array}{ll}
		\bot  & \quad \mbox{if } C \in A\\
		\neg C|_{AB} & \quad \mbox{if } C \in B\\
	\end{array}
\right. $
}\\
\hline
\end{tabular}
\end{minipage}
\qquad
\begin{minipage}[t]{.60\textwidth}
\begin{tabular}{|l c|c|c|c|}%
\hline
Inner node: & \multicolumn{4}{c|}{$\quad \dfrac{C_1 \vee p \, [I_1] \qquad  C_2 \vee \n{p} \, [I_2]}{C_1 \vee C_2 \, [I]}$} \\
\hline
\multicolumn{5}{|c|}{$I =
\left\{
	\begin{array}{ll}
		I_1 \vee I_2  & \quad \mbox{if } p \in A \text{ or } p \in AB \\
		I_1 \wedge I_2 & \quad \mbox{if } p \in B
	\end{array}
\right. $
}\\
\hline
\end{tabular}
\end{minipage}
\caption{McMillan$'$ interpolation algorithm.}
\label{fig:mcmp}
\end{figure}

Alg.~\ref{alg:reor2} can be employed to transform the refutation; in this case  a context is unordered if $v(t)$ is a variable local to $B$ (\emph{light}) and $v(s)$ is a
variable local to $A$ or shared (\emph{heavy}). The effect of pushing up light variables is that, during the construction of the interpolant,
the connective $\wedge$ will be introduced before $\vee$ along each path, so that the resulting interpolant will be in disjunctive normal form (note that the partial interpolant of
a leaf is already in DNF, being a conjunction of literals).

We can thus state the following theorem:
\begin{thm}
\label{thm:dnfization}
Assume a refutation $P$ without unordered contexts.  McMillan$'$ interpolation algorithm generates an interpolant in DNF from $P$.
\end{thm}
As already pointed out in~\cite{JM05}, the price to pay for a complete transformation might be an exponential increase of the proof size,
due to the node duplications necessary to apply rules $S1,S2,R2$ to contexts where $C_4$ has multiple children (see Fig.~\ref{fig:rulesdag}).
A feasible compromise consists in performing a partial CNFization or DNFization by limiting the application of such rules to 
when $C_4$ has a single child; in this case, the proof growth depends only on the application of rule $S1$, and the increase is maintained linear.


\section{Heuristics for the Proof Transformation Algorithms}
\label{sec:heuri}

In this section we discuss some of the heuristics implemented in \opensmt and \periplo to guide the application of the
Local Transformation Framework rules and the reconstruction of proofs, 
with reference to compression (\S\ref{sec:compr}) and pivot reordering for interpolation (\S\ref{sec:reordtrasf}).

Some of the algorithms presented so far (Algs.~\ref{alg:prop},\ref{alg:rectraloop},\ref{alg:simp}) 
need to handle the presence of resolution steps which are not valid anymore since the pivot is missing from both antecedents;
in that case, the resolvent node $n$ must be replaced by either parent. A heuristics which has been proven useful for determining
the replacing parent is the following. If one of the parents (let us say $n^+$) has only $n$ as child, then  
$n$ is replaced by $n^-$; since $n^+$ loses its only child, then (part of) the subproof rooted in $n^+$ gets detached from the global proof,
yielding a simplification of the proof itself. If both parents have more than one child, then the parent labeled by the smaller
clause is the one that replaces $n$, aiming at increasing the amount of simplifications performed while moving down to the global root.

As far as the heuristics for the application of rewriting rules are concerned, the ApplyRule method adheres to some general lines.
Whenever a choice is possible between a left and a right context, a precedence order is respected:
($X > Y$ means: the application of $X$ is preferred over that of $Y$):
\[
R3 > \{ R2', R1\} > R2 > S1' > S2 > S1
\]
The compression rules $R$ have always priority over the shuffling rules $S$, $R3$ being the favorite, followed by $R2'$ and $R1$.
Among the $S$ rules, $S1'$ is able to perform a local simplification, which makes it preferred to $S2$ and especially to $S1$, which
increases the size of the proof; between equal $S$ rules, the one which does not involve a node duplication (see Fig.~\ref{fig:rulesdag})
is chosen.

Additional constraints depend on the actual goal of the transformation. 
If the aim is pivot reordering, the constraints are as illustrated in Alg.~\ref{alg:reor2},
with ties broken according to the general lines given above.
If the aim is compression, then $S1$ is never applied, since it increases 
the size of the proof and it is not apparent at the time of its application whether it would bring benefits in a second moment, neither are applied
$R2, S1',S2$ if they involve a duplication. A strategy which proved successful in the application of $S$ rules is to push up
nodes with multiple resolvents whenever possible, with the aim of improving the effect of RecyclePivots and RecyclePivotsWithIntersection; interestingly, this technique
shows as a side effect the disclosure of redundancies which can effectively be taken care of by StructuralHashing.  

These heuristics have been discovered through experimentation and have been adopted due to their practical usefulness for compression,
in a setting where the large size of proofs allows only a few traversals (and thus a limited application of rules) by means of ReduceAndExpose,
and where the creation of new nodes should be avoided; it is thus unlikely that, arbitrarily increasing the number of traversals, they would
expose and remove all pivots redundancies. A more thorough, although practically infeasible, approach could rely
on keeping track of all contexts and associated rules in a proof $P$. Since the $S$ rules are revertible, an equivalence relation $\equiv_S$ could
be defined among proofs so that $P\equiv_S P'$ if $P'$ can be obtained from $P$ (and vice versa) by means of a sequence of applications of $S$ rules.   
A backtracking-based algorithm could be employed to systematically visit equivalence classes of proofs, and to move from an equivalence class to another
thanks to the application of an $R$ rule.

\section{Related Work}
\label{sec:relwork}

Various proof manipulation techniques have been developed in the last years,
the main goal being compression. 

In~\cite{A06}, Amjad proposes an algorithm based on heuristically reordering the resolution steps that form a proof,
trying to identify a subset of the resolution steps that is still sufficient to derive the empty clause. The approach
relies on using an additional graph-like data structure to keep track of how literals of opposite polarity are propagated
from the leaves through the proof and then resolved upon.

Sinz~\cite{S07} explicitly assumes a CDCL context, where a resolution-based SAT-solver generates a sequence of
derivations called \emph{proof chains}, combined in a second moment to create the overall proof.
He presents an algorithm that works at the level of proof chains, aiming at identifying and merging shared
substructures to generate a smaller proof.

Amjad further develops this approach in~\cite{A08}. He adopts a representation of resolution proofs that allows the use
of efficient algorithms and data structures for substring matching; this feature is exploited to perform memoization of proofs by
detecting and reusing common subproofs. 

Cotton introduces in~\cite{C10} two compression methods. 
The first one is based on a form of structural hashing, where each inner node in a proof graph is associated with its pair of antecedents in a hash map.
The compression algorithm traverses the sequence of proof chains while updating the hash map, and adds a resolution step to the overall proof only if 
it does not already exist.
The second one consists of a rewriting procedure that, given in input a proof and a heuristically chosen propositional variable $p$, transforms the proof so that the 
last resolution step is on $p$; this might result in a smaller proof.

Bar-Ilan et al.~\cite{BFHSS08} present a technique that exploits learned unit clauses to rewrite subproofs  
that were derived before learning them. They also propose a compression algorithm (\emph{RecyclePivots}) that searches for resolution steps on the same pivot
along paths from leaves to the root in a proof. If a pivot is resolved upon more than once on a path (which implies that 
the pivot variable is introduced and then removed multiple times), the resolution step closest to the root is kept, 
while the others are simplified away. The algorithm is effective on resolution proof trees, but can be applied only in a limited
form to resolution proof DAGs, due to the possible presence of multiple paths from a node to the root.

This restriction is relaxed in the work of Fontaine et al.~\cite{FMP11}, who extend the algorithm of~\cite{BFHSS08} into \emph{RecyclePivotsWithIntersection} to keep track, for each node, 
of the literals which get resolved upon along \emph{all} paths from the node to the root. 
\cite{FMP11} also presents an algorithm that traverses a proof, collecting unit clauses and reinserting them  
at the level of the global root, thus removing redundancies due to multiple resolution steps on the same unit clauses;
this technique is later generalized in~\cite{BPa13} to lowering subproofs rooted in non-unit clauses.

\cite{Ag12} builds upon~\cite{BFHSS08} in developing three variants of \emph{RecyclePivots} tailored to resolution proof DAGs.
The first one is based on the observation that the set of literals which get resolved in a proof upon along all paths from the node to the root
must be a superset of the clause associated to the node, if the root corresponds to the empty clause.
The second and third ones actually correspond respectively to \emph{RecyclePivotsWithIntersection} and to a parametric version of it
where the computation of the set of literals is limited to nodes with up to a certain amount of children. 

Our set of compression techniques has been illustrated with reference to~\cite{S07}, \cite{BFHSS08} and \cite{FMP11} in \S\ref{sec:compr}.  

Besides compression, a second area of application of proof manipulation has been interpolation, both in the propositional and in the first order settings.

D'Silva et al.~\cite{DKPW08} introduce a global transformation framework for interpolation to reorder the resolution steps in a proof with respect to a 
given partial order among pivots; compression is shown to be a side effect for some benchmarks.
Compared to~\cite{DKPW08}, our approach
works locally, and leaves more freedom in choosing the 
strategies for rule applications. Also our target is not directly
 computing interpolants, but rather rewriting the proof in such
a way that existing techniques can be applied. 

The same authors focus in~\cite{DKPW10} on the concept of 
strength of an interpolant. They present an analysis 
of existing propositional interpolation algorithms, together with a method to combine 
them in order to obtain weaker or stronger interpolants from a same 
proof of unsatisfiability. They also address the use and the limitations
of the local transformation rules of Jhala and McMillan~\cite{JM05}.
The rewriting rules corresponding to $S1$ and $S2$ in the Local Transformation Framework (\S\ref{sec:tra_fra}) were first 
introduced in~\cite{JM05} and further examined in~\cite{DKPW10} as a way to modify a proof to obtain stronger or weaker interpolants,
once fixed the interpolation algorithm; we devised the remaining rules after an exhaustive analysis 
of the possible proof contexts. 
\cite{JM05} also discusses the application of $S1$ and $S2$ to  generate interpolants in conjunctive normal form;
however, not all the contexts are taken into account, and, as pointed out in~\cite{DKPW10}, the contexts for  $S1$ and $S2$
are not correctly identified.

Note that $S1$ and $S2$ have also a counterpart in Gentzen's sequent calculus system $LK$~\cite{gentz}:
$S1$ corresponds to swapping applications of the structural cut and contraction rules,
while $S2$ is one of the rank reduction rules.

Interpolation for first order theories in presence of $AB$-mixed predicates is addressed in~\cite{CGS08}, only for the case of DTC,
by tweaking the decision heuristics of the solver, in such a way that
it guarantees that the produced proof can be handled with known methods.
In particular the authors define a notion of $ie$-\emph{local proofs}, and they
show how to compute interpolants for this class of proofs,
and how to adapt an SMT-solver to produce only $ie$-local
proofs. \cite{GKT09} the relaxes the constraint on
generating $ie$-local proofs by introducing the notion of 
{\em almost-colorable} proofs. We argue that our technique
is simpler and more flexible, as different strategies can
be derived with different applications of our local transformation rules. 
Our method is also more general, since it applies not 
only to theory combination but to any approach that requires the addition of
$AB$-mixed predicates (see \S\ref{subsec:applications}).

More recently, a tailored interpolation algorithm has been proposed in~\cite{CHN13} for the combined theory
of linear arithmetic and uninterpreted functions; it has the notable feature of allowing the presence of mixed predicates,
thus making proof manipulation not necessary anymore.

\paragraph{Clausal Proofs.}
This paper addresses resolution proofs in the context of transformation for compression and Craig interpolation;
state-of-the-art algorithms, as described in the previous sections, rely on representing and manipulating proofs in the form of directed acyclic graphs.
However, alternative approaches exist; for example, CDCL SAT-solvers can be instrumented to generate  proofs in \emph{clausal format}, as a sequence of
learned clauses~\cite{GN03,Gel08,HHW13}. The development of compression techniques tailored to clausal proofs is an interesting topic, which will be investigated
as future work.


\section{Conclusions}
\label{sec:concl}

In this paper we have presented a proof transformation framework based on a
set of local rewriting rules and shown how it can be applied to the tasks of proof compression
and pivot reordering.

As for compression, we discussed how rules that effectively simplify the proof can be interleaved
with rules that locally perturbate the topology, in order to create new opportunities for compression.
We identified two kinds of redundancies in proofs, related to the notions of regularity and compactness,
and presented and compared a number of algorithms to address them, moving from existing techniques in the literature.
Individual algorithms, as well as their combinations, were implemented and tested over a collection of benchmarks both from SAT and SMT
libraries, showing remarkable levels of compression in the proof size.

As for pivot reordering, we described how to employ the rewriting rules to isolate and remove
$AB$-mixed predicates, in such a way that standard procedures for interpolation in SMT
can be applied. The approach
enables the use of off-the-shelf techniques for
SMT-solvers that are likely to introduce $AB$-mixed
predicates, such as Ackermann's Expansion, Lemma on Demand, 
Splitting on Demand and DTC. We showed by means of experiments
that our rules can effectively transform the proofs without
generating any exponential growth in their size.
Finally, we explored a form of interaction between LISs and proof manipulation by providing algorithms to reorder resolution steps in a propositional proof
to guarantee the generation of interpolants in conjunctive or disjunctive normal form.





\end{document}